\def\year{2021}\relax
\documentclass[letterpaper]{article} %
\usepackage{aaai21}  %
\usepackage{times}  %
\usepackage{helvet} %
\usepackage{courier}  %
\usepackage[hyphens]{url}  %
\usepackage{graphicx} %
\usepackage{natbib}  %
\usepackage{caption} %
\usepackage{subfigure}
\usepackage{amsmath}
\usepackage{multirow}
\usepackage{array}
\usepackage{booktabs}
\usepackage{tabularx}
\usepackage[switch]{lineno}  %
\usepackage{fancyvrb}
\usepackage{xcolor}


\newcommand{\CommentA}{$C_{1}$}
\newcommand{\CommentB}{$C_{2}$}

\newcommand{\CodeA}{$M_{1}$}
\newcommand{\CodeB}{$M_{2}$}

\newcommand{\CodeIn}[1]{{\ifmmode{\mathtt{#1}}\else$\mathtt{#1}$\fi}}

\newcommand{\Return}{\CodeIn{@return}}
\newcommand{\Param}{\CodeIn{@param}}
\newcommand{\Comment}{$C$}

\newcommand{\OldCode}{$M_{old}$}
\newcommand{\NewCode}{$M$}

\newcommand{\NewComment}{$C_{new}$}
\newcommand{\EditComment}{$C_{edit}$}
\newcommand{\Posthoc}{post hoc}
\newcommand{\JustInTime}{just-in-time}
\newcommand{\EditCode}{$M_{edit}$}
\newcommand{\NewTree}{$T$}
\newcommand{\OldTree}{$T_{old}$}
\newcommand{\EditTree}{$T_{edit}$}
\newcommand{\SeqModel}{\textsc{Seq}}
\newcommand{\GraphModel}{\textsc{Graph}}
\newcommand{\HybridModel}{\textsc{Hybrid}}

\newcommand{\HasOverlap}{\textsc{Overlap}}
\newcommand{\DeletedCode}{deleted}

\newcommand{\PosthocSeq}{\SeqModel{}(\Comment{}, \NewCode{})}
\newcommand{\PosthocGraph}{\GraphModel{}(\Comment{}, \NewTree{})}
\newcommand{\PosthocHybrid}{\HybridModel{}(\Comment{}, \NewCode{}, \NewTree{})}
\newcommand{\JustInTimeSeq}{\SeqModel{}(\Comment{}, \EditCode{})}
\newcommand{\JustInTimeGraph}{\GraphModel{}(\Comment{}, \EditTree{})}
\newcommand{\JustInTimeHybrid}{\HybridModel{}(\Comment{}, \EditCode{}, \EditTree{})}

\newcommand{\UpdateCopy}{Update w/ implicit detection}
\newcommand{\Pretrained}{Pretrained update + detection}
\newcommand{\JointlyTrained}{Jointly trained update + detection}
\newcommand{\GenMatch}{\textsc{GenMatch}}

\newcommand{\JustInTimeSeqImplicit}{\SeqModel{}(\Comment{}, \OldCode{}, \NewCode{})}
\newcommand{\JustInTimeGraphImplicit}{\GraphModel{}(\Comment{}, \OldTree{}, \NewTree{})}
\newcommand{\JustInTimeHybridImplicit}{\HybridModel{}(\Comment{}, \OldCode{}, \NewCode{}, \OldTree{}, \NewTree{})}

\newcommand{\Bert}{CodeBERT BOW}
\newcommand{\PosthocBert}{\Bert{}(\Comment{}, \NewCode{})}
\newcommand{\JustInTimeBert}{\Bert{}(\Comment{}, \EditCode{})}

\def\@fnsymbol#1{\ensuremath{\ifcase#1\or *\or \dagger\or \ddagger\or
   \mathsection\or \mathparagraph\or \|\or **\or \dagger\dagger
   \or \ddagger\ddagger \else\@ctrerr\fi}}
\newcommand{\ssymbol}[1]{^{\@fnsymbol{#1}}}

\newcommand{\NumOfProjects}{1,518}
\newcommand{\NumOfExamples}{40,688}

  

\title{Deep Just-In-Time Inconsistency Detection Between Comments and Source Code}

\author{
Sheena Panthaplackel\textsuperscript{\rm 1},
Junyi Jessy Li\textsuperscript{\rm 2},
Milos Gligoric\textsuperscript{\rm 3},
Raymond J. Mooney\textsuperscript{\rm 1}\\
}

\affiliations{
\textsuperscript{\rm 1}Department of Computer Science\\
\textsuperscript{\rm 2}Department of Linguistics\\
\textsuperscript{\rm 3}Department of Electrical and Computer Engineering\\
The University of Texas at Austin\\
spantha@cs.utexas.edu, jessy@austin.utexas.edu, gligoric@utexas.edu, mooney@cs.utexas.edu
}

\begin{document}

\maketitle

\begin{abstract}
Natural language comments convey key aspects of source code such as implementation, usage, and pre- and post-conditions. Failure to update comments accordingly when the corresponding code is modified introduces inconsistencies, which is known to lead to confusion and software bugs. In this paper, we aim to detect whether a comment becomes inconsistent as a result of changes to the corresponding body of code, in order to catch potential inconsistencies \emph{just-in-time}, i.e., before they are committed to a code base. To achieve this, we develop a deep-learning approach that learns to correlate a comment with code changes. By evaluating on a large corpus of comment/code pairs spanning various comment types, we show that our model outperforms multiple baselines by significant margins. For extrinsic evaluation, we show the usefulness of our approach by combining it with a comment update model to build a more comprehensive automatic comment maintenance system which can both detect and resolve inconsistent comments based on code changes.
\end{abstract}

\section{Introduction}
Comments serve as a critical communication medium for developers, facilitating program comprehension and code maintenance tasks~\cite{CodeReadabilityBuse,deSouzaMaintenance}. Code is highly-dynamic in nature, with developers constantly making changes to address bugs and feature requests. Many code changes require reciprocal updates to the accompanying comments to keep them in sync; however, this is not always done in practice~\cite{WenLargeStudy, FluriAnalysis,ratol2017fragile, JiangEvolution, ZhouParameter,icomment2007}. Outdated comments which inaccurately portray the code they accompany adversely affect the software development cycle by causing confusion~\cite{WenLargeStudy, JiangEvolution, icomment2007, ZhouParameter} and misguiding developers, hence making code vulnerable to bugs~\cite{JiangEvolution, icomment2007,IbrahimBugs}. Therefore, it is desirable to have systems that can automatically detect such inconsistencies and alert developers.

Previous work has explored heuristic-based approaches for automatically detecting specific types of inconsistencies
(e.g., identifier naming~\cite{ratol2017fragile}, parameter constraints~\cite{ZhouParameter}, \texttt{null} values and exceptions~\cite{tComment}, locking~\cite{icomment2007}, interrupts~\cite{aComment}). Some have also addressed the notion of coherence
between comments and code as a text similarity problem with traditional machine learning models that leverage bag-of-words techniques~\cite{Corazza18,Cimasa19}. In contrast, we design an approach that generalizes across types
of inconsistencies
and captures deeper comment/code relationships. Furthermore, prior research has predominantly focused on detecting inconsistencies that already reside in a software project, within the code repository. We refer to this as \textit{\Posthoc{} inconsistency detection} since it occurs potentially many commits \textit{after} the inconsistency has been introduced. %

Ideally, these inconsistencies should be detected before they ever enter the repository (e.g., during code review) since they pose a threat to the development cycle and reliability of the software until they are found. Because inconsistent comments generally arise as a consequence of developers failing to update comments immediately following code changes~\cite{WenLargeStudy}, we aim to detect whether a comment becomes inconsistent as a result of changes to the accompanying code, \textit{before} these changes are merged into a code base. We refer to this as \textit{\JustInTime{} inconsistency detection}, as it allows catching potential inconsistencies right before they can materialize.

\begin{figure}[t]
\centering
\subfigure[Inconsistent]{
    \label{fig:apache-ignite}
    \includegraphics[scale=0.25]{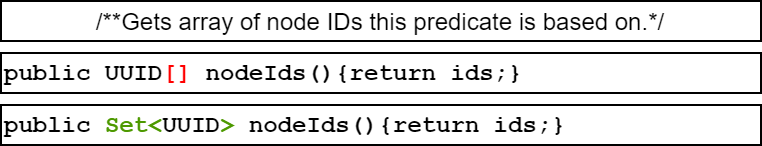}
}
\subfigure[Consistent]{
    \label{fig:alluxio}
    \includegraphics[scale=0.25]{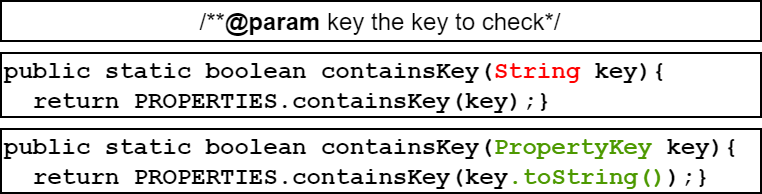}
}
\vspace{-5pt}
\caption{In the example from the Apache Ignite project shown in Figure~\ref{fig:apache-ignite}, the existing comment becomes inconsistent upon changes to the corresponding method, and in the example from the Alluxio project shown in Figure~\ref{fig:alluxio}, the existing comment remains consistent after code changes.}
\label{fig:intro_example}
\end{figure}

Detecting inconsistencies immediately following code changes allows us to utilize information from the version of the code before the changes, for which the comment is consistent. By considering how the changes affect the relationship the comment holds with the code, we can determine whether the comment remains consistent after the changes. For instance, in Figure~\ref{fig:apache-ignite}, the comment describes the return type of \CodeIn{nodeIds()} as an array. When the method is modified to return a \CodeIn{Set} instead of an array, the comment no longer describes the correct return type, making it inconsistent. Such analysis is not possible in \Posthoc{} inconsistency detection since the exact code changes that triggered inconsistency cannot be easily pinpointed, making it difficult to align the comment with relevant parts of the code. 

Moreover, due to challenges in crafting  data extraction rules~\cite{icomment2007,aComment} and annotating substantial amounts of data~\cite{Corazza18}, prior \Posthoc{} work relies on a limited set of examples and projects. In contrast, we build a large corpus for \JustInTime{} inconsistency detection by mining commit histories of software projects for  code changes with and without corresponding comment updates.

Few approaches exploit code changes for inconsistency detection and these rely on task-specific rules~\cite{SaduThesis}, hand-engineered surface features~\cite{LiuOutdatedLine,Malik08}, and bag-of-words techniques~\cite{LiuOutdatedLine}.
Instead, we  \textit{learn} salient characteristics of these inputs through a deep-learning framework that encodes their syntactic structures. Namely, we use recurrent neural networks (RNNs) and gated graph neural networks (GGNNs)~\cite{Li2016GatedGS} to learn contextualized representations of the comment and code changes and multi-head attention~\cite{transformer}
to relate these representations in order to discern how the code changes affect the comment. We also study how manual features can complement our neural approach.

Furthermore, on its own,
an inconsistency detection system can only flag comments that developers failed to update. Actually amending them to reflect code changes requires significant developer effort. Approaches for automatically updating comments based on code changes have been recently proposed~\cite{panthaplackel2020update, LiuJITUpdate}. However, they do not handle cases in which an update is not needed, such as in 
Figure~\ref{fig:alluxio}. While the type of the \CodeIn{key} argument is modified, its purpose is unchanged (i.e., it still represents the key to be checked in \CodeIn{PROPERTIES}). Based on our user study~\cite{panthaplackel2020update}, such cases deteriorated the overall quality of the system.
As a form of extrinsic evaluation, we evaluate the utility of our approach by integrating
it with this comment update model, to build a more comprehensive automatic comment maintenance system that detects and resolves inconsistencies.

To summarize, our main contributions are as follows: (1)~We develop a deep learning approach for \JustInTime{} inconsistency detection that correlates a comment with changes in the corresponding body of code and which outperforms the \Posthoc{} setting as well as several baselines. (2)~For training and evaluation, we construct a large corpus of comments paired with code changes in the corresponding methods, encompassing multiple types of method comments and consisting of \NumOfExamples{} examples that are extracted from \NumOfProjects{} open-source Java projects.\footnote{Data and implementation are available at \url{https://github.com/panthap2/deep-jit-inconsistency-detection}.} (3)~We demonstrate the value of inconsistency detection in a comprehensive automatic comment maintenance system, and we show how our approach can support such a system.

\section{Task}

Our task is to determine whether a comment is inconsistent, or semantically out of sync with the corresponding method. Most inconsistencies result from developers making code changes without properly updating the accompanying comments. Suppose \OldCode{} from the consistent comment/method pair (\Comment{}, \OldCode{}) is modified to \NewCode{}. If \Comment{} is not in sync with \NewCode{} and is not updated, it will become inconsistent once \NewCode{} is committed. We frame this problem in two distinct settings, with the task being constant across both: determine whether \Comment{} is inconsistent with \NewCode{}.

\begin{itemize}
  \item \textbf{Post hoc:} Here, only the existing version of the comment/method pair is available; the code changes that triggered the inconsistency are unknown. 
  \item \textbf{Just-in-time:} Here, the goal is to catch inconsistencies before they are committed. Unlike the \Posthoc{} setting, \OldCode{} is available, allowing us to analyze the changes between \OldCode{} and \NewCode{}.
\end{itemize}

In line with most prior work in inconsistency detection~\cite{Corazza18, icomment2007,tComment,Khamis2010AutomaticQA}, we focus on identifying inconsistencies in comments comprising API documentation for Java methods. API documentation consists of a main description and a set of tag comments~\cite{javadoc}. While some have considered treating the full documentation as a single comment~\cite{Corazza18}, we choose to perform inconsistency detection at a more fine-grained level, analyzing individual comment types within this documentation. Furthermore, in contrast to previous studies tailored to a specific tag~\cite{ZhouParameter, tComment} or specific keywords and templates~\cite{icomment2007,aComment}, we simultaneously consider multiple comment types with diverse characteristics. Namely, we address inconsistencies in the \Return{} tag comment, which describes a method's return type, and the \Param{} tag comment, which describes an argument of the method. Additionally, we examine inconsistencies in the less-structured summary comment, derived from the first sentence of the main description.

\begin{figure}
\centering
\includegraphics[width=\columnwidth]{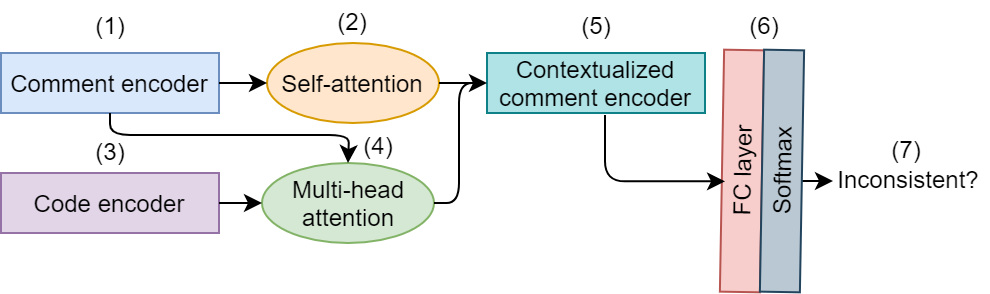}
\vspace{-5pt}
\caption{High-level architecture of our approach.}
\label{fig:architecture}
\end{figure}

\section{Architecture}
We aim to determine whether \Comment{} is inconsistent by understanding its semantics
and how it relates to \NewCode{} (or changes between \OldCode{} and \NewCode{}). We show an overview of our approach in Figure~\ref{fig:architecture}. First, the comment encoder, a BiGRU~\cite{ChoGRU}, encodes the sequence of tokens in \Comment{} (Figure~\ref{fig:architecture} (1)).
When learning a representation for a given token, the forward and backward BiGRU passes, in principle, provide context of other tokens in \Comment{}. However, this information can get diluted, especially when there are long-range dependencies, and the relevant context can also vary across tokens. To address this, we update these representations from the comment encoder with more context about how they relate to the other tokens through multi-head self-attention~\cite{transformer} (Figure~\ref{fig:architecture} (2)). Next, we learn code representations with a code encoder (Figure~\ref{fig:architecture} (3)), which can be a sequence encoder (cf. \S\ref{sec:sequence-code-encoder}) or an abstract syntax tree (AST) encoder (cf. \S\ref{sec:ast-code-encoder}).

Since the essence of the task comes down to whether \Comment{} accurately reflects \NewCode{}, we must capture the relationship between \Comment{} and \NewCode{} (or changes between \OldCode{} and \NewCode{}). Prior work does this by computing comment/code similarity through lexical overlap rules~\cite{ratol2017fragile,SaduThesis}, which do not work well when different terms have similar meanings, and cosine similarity between vector representations, which have been found to perform poorly on their own~\cite{LiuOutdatedLine,Cimasa19}. Furthermore, this notion of similarity is only appropriate for the summary comment which provides an overview of the corresponding method as a whole. More specialized comment types like \Return{} and \Param{} describe only specific parts of the method. Therefore, their representations may not be very similar to the representation of the full method. In contrast, we learn the relationship between comments and code by computing multi-head attention between each hidden state of the comment encoder and the hidden states of the code encoder (Figure~\ref{fig:architecture} (4)).

\begin{figure}
\centering
\includegraphics[width=\columnwidth]{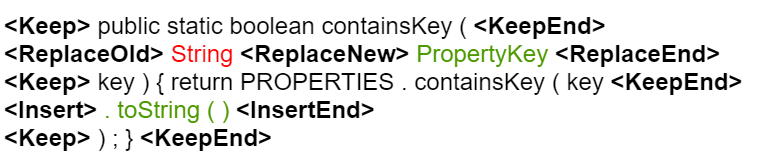}
\vspace{-5pt}
\caption{Sequence-based code edit representation (\EditCode{}) corresponding to Figure~\ref{fig:alluxio}, with removed tokens in red and added tokens in green.}
\label{fig:diff_sequence}
\end{figure}

\begin{figure}
\centering
\includegraphics[width=\columnwidth]{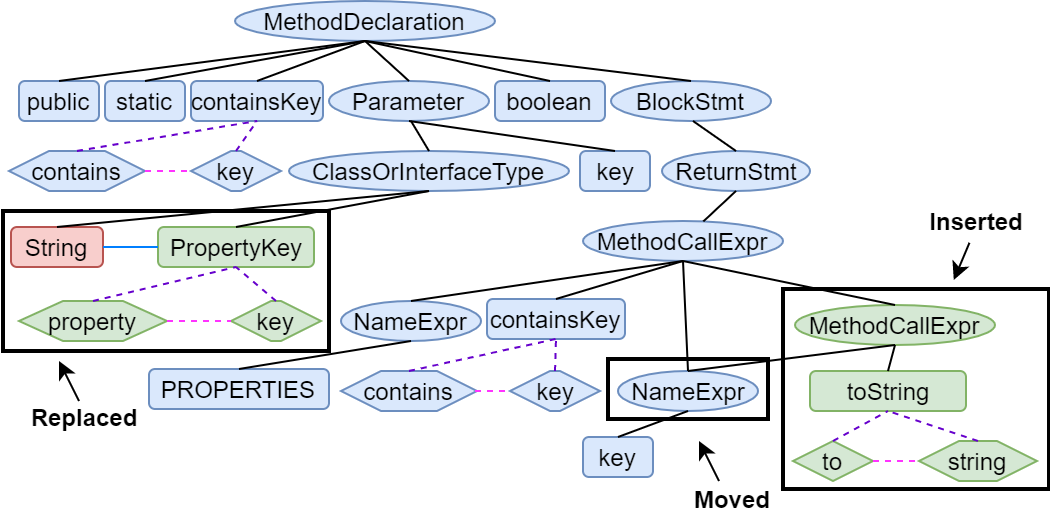}
\vspace{-5pt}
\caption{AST-based code edit representation (\EditCode{}) corresponding to Figure~\ref{fig:alluxio}, with removed nodes in red and added nodes in green.}
\label{fig:diff_ast}
\vspace{-10pt}
\end{figure}

We combine the context vectors resulting from both attention modules to form enhanced representations of the tokens in \Comment{}, which carry context from other parts of \Comment{} as well as the code. These are then passed through another BiGRU encoder (Figure~\ref{fig:architecture} (5)). We take the final state of this encoder to be the vector representation of the full comment, and we feed it through fully-connected and softmax layers (Figure~\ref{fig:architecture} (6)). This leads to the final prediction (Figure~\ref{fig:architecture} (7)).

\subsection{Sequence Code Encoder}
\label{sec:sequence-code-encoder}
In the \JustInTime{} setting, we represent the changes between \OldCode{} and \NewCode{} with an edit action sequence, \EditCode{}. We have previously shown that explicitly defining edits in such a way outperforms having the model implicitly learn them~\cite{panthaplackel2020update}. Each action consists of an action type (\CodeIn{Insert}, \CodeIn{Delete}, \CodeIn{Keep}, \CodeIn{ReplaceOld}, \CodeIn{ReplaceNew}) that applies to a span of tokens, as shown in Figure~\ref{fig:diff_sequence}. We encode \EditCode{} with a BiGRU. Because \OldCode{} is unavailable in the \Posthoc{} setting, we cannot construct an edit action sequence. So, we encode the sequence of tokens in \NewCode{}.

\subsection{AST Code Encoder}
\label{sec:ast-code-encoder}
To better exploit the syntactic structure of code, we leverage its abstract syntax tree (AST). Following prior work in other tasks~\cite{FernandesSummarization, yin19iclr}, we encode ASTs and AST edits using gated graph neural networks (GGNNs)~\cite{Li2016GatedGS}. For the \Posthoc{} setting, we encode \NewTree{}, an AST-based representation corresponding to \NewCode{}. In the \JustInTime{} setting, we instead encode \EditTree{}, an AST-based edit representation. We use GumTree \cite{GumTree}, to compute AST node edits between \OldTree{} (corresponding to \OldCode{}) and \NewTree{}, identifying inserted, deleted, kept, replaced, and moved nodes. We merge the two, forming a unified representation, by consolidating identical nodes, as shown in Figure~\ref{fig:diff_ast}.

GGNN encoders for \NewTree{} and \EditTree{} use \textit{parent} (\CodeIn{public} $\rightarrow$ \CodeIn{MethodDeclaration}) and \textit{child} (\CodeIn{MethodDeclaration} $\rightarrow$ \CodeIn{public}) edges. Like prior work~\cite{FernandesSummarization}, we add ``subtoken nodes" for identifier leaf nodes to better handle previously unseen identifier names. To integrate these new nodes, we add \textit{subnode} (\CodeIn{toString} $\rightarrow$ \CodeIn{to}), \textit{supernode} (\CodeIn{to} $\rightarrow$ \CodeIn{toString}), \textit{next subnode} (\CodeIn{to} $\rightarrow$ \CodeIn{string}), and \textit{previous subnode} (\CodeIn{string} $\rightarrow$ \CodeIn{to}) edges. When encoding \EditTree{}, we also include an \textit{aligned} edge type between nodes in the two trees that correspond to an update (\CodeIn{String} and \CodeIn{PropertyKey}). Additionally, we learn \textit{edit} embeddings for each action type. To identify how a node is edited (or not edited), we concatenate the corresponding edit embedding to its initial representation that is fed to the GGNN.

\section{Data}
\label{sec:data}
By detecting inconsistencies at the time of code change, we can extract automatic supervision from commit histories of open-source Java projects. Namely, we compare consecutive commits, collecting instances in which a method is modified. We extract the comment/method pairs from each version: (\CommentA{}, \CodeA{}), (\CommentB{}, \CodeB{}). In prior work, we isolate comment updates made based on code changes through cases in which \CommentA{}$\neq$\CommentB{}~\cite{panthaplackel2020update}. By assuming that the developer updated the comment because it would have otherwise become inconsistent as a result of code changes, we take \CommentA{} to be inconsistent with \CodeB{}, consequently leading to a \emph{positive example}, with \Comment{}=\CommentA{}, \OldCode{}=\CodeA{}, and \NewCode{}=\CodeB{}. For \emph{negative examples}, we additionally examine cases in which \CommentA{}=\CommentB{} and assume that if the existing comment would have become inconsistent, the developer would have updated it.
Following this process, we collect \Return{}, \Param{}, and summary comment examples. We additionally incorporate 7,239 positive \Return{} examples from our prior work~\cite{panthaplackel2020update} which studies \Return{} comment updates.

\begin{table}
\centering
\small
\begin{tabular}{l@{\hskip 3mm}rrrr}
\hline
&  \bf Train & \bf Valid & \bf Test & \bf Total\\
\hline
@return & 15,950 & 1,790 & 1,840 & 19,580\\
@param & 8,640 & 932 & 1,038 & 10,610\\
Summary & 8,398 &  1,034 & 1,066 & 10,498\\
\hline
Full & 32,988 & 3,756 & 3,944 & 40,688\\
\hline
Projects & 829 & 332 & 357 & 1,518\\
\hline
\end{tabular}
\vspace{-5pt}
\caption{\label{table:data-splits}Data partitions.}
\end{table}

While convenient for data collection, the assumptions we make do not always hold in practice. For instance, if \CommentA{} is refactored without altering its meaning, we would assign a positive label because \CommentA{}$\neq$\CommentB{}, despite it actually being consistent. Because such cases of \textit{comment improvement} are not within the scope of our work, we adopt previously proposed heuristics~\cite{panthaplackel2020update} to reduce the number of instances in which the comment and code changes are unrelated. The negative label is also noisy since \CommentA{}=\CommentB{} when a developer fails to update comments in accordance with code changes, pointing to the problem we are addressing in this paper. We minimize such cases by limiting to popular, well-maintained projects~\cite{ProjectQuality}. For more reliable evaluation, we curate a clean sample of 300 examples
(corresponding to 101 projects) from the test set, consisting of 50 positive and 50 negative examples of each comment type.

In line with prior work~\cite{RenCrossProject,Movshovitz-AttiasCohen13PredictingProgrammingComments}, we consider a cross-project setting with no overlap between the projects from which examples are extracted in training/validation/test sets. From our data collection procedure, we obtain substantially more negative examples than positive ones, which is not surprising because many changes do not require comment updates~\cite{WenLargeStudy}. We downsample negative examples, for each partition and comment type, to construct a balanced dataset.
Statistics of our final dataset are shown in Table~\ref{table:data-splits}.

Comments are tokenized based on space and punctuation. We parse methods into sequences using javalang~\cite{javalang}. Comment and code sequences are subtokenized (e.g., camelCase $\rightarrow$ camel, case; snake\_case $\rightarrow$ snake, case), as done in prior
work~\cite{Alon2019Code2Seq,FernandesSummarization}, to capitalize on composability and better address the open vocabulary problem in learning from source code~\cite{CvitovicOpenVocab}. 
Details on data statistics, filtering, and annotation procedures are given in Appendix~\ref{appendix:data}.

\section{Models}
We outline baseline, \Posthoc{}, and \JustInTime{} inconsistency detection models.

\subsection{Baselines}
\label{sec:baselines}
\noindent\textbf{Lexical overlap:} A comment often has lexical overlap with the corresponding method. We include a rule-based \JustInTime{} baseline, \HasOverlap{}(\Comment{}, \DeletedCode{}), which classifies \Comment{} as inconsistent if at least one of its tokens matches a code token belonging to a \CodeIn{Delete} or \CodeIn{ReplaceOld} span in \EditCode{}.

\noindent\textbf{\citeauthor{Corazza18}~\shortcite{Corazza18}:} This \Posthoc{} bag-of-words approach classifies whether a comment is coherent with the method that it accompanies using an SVM with TF-IDF vectors corresponding to the comment and method. We simplify the original data pre-processing, but validate that the performance matches the reported numbers.

\noindent\textbf{\Bert{}:} We develop a more sophisticated bag-of-words baseline that leverages CodeBERT~\cite{Feng2020CodeBERTAP} embeddings. These embeddings were pretrained on a large corpus of natural language/code pairs. In the \Posthoc{} setting, we consider \PosthocBert{}, which computes the average embedding vectors of \Comment{} and \NewCode{}. These vectors are concatenated and fed through a feedforward network. In the \JustInTime{} setting, we compute the average embedding vector of \EditCode{} rather than \NewCode{}, and we refer to this baseline as \JustInTimeBert{}.

\noindent\textbf{\citeauthor{LiuOutdatedLine}~\shortcite{LiuOutdatedLine}:} This is a \JustInTime{} approach for detecting whether a block/line comment becomes inconsistent upon changes to the corresponding code snippet. Their task is slightly different as block/line comments describe low-level implementation details and generally pertain to only a limited number of lines of code, relative to API comments. However, we consider it as a baseline since it is closely related. They propose a random forest classifier which leverages features which capture aspects of the code changes (e.g., whether there is a change to a \CodeIn{while} statement), the comment (e.g., number of tokens), and the relationship between the comment and code (e.g., cosine similarity between representations in a shared vector space). We re-implemented this approach based on specifications in the paper, as their code was not publicly available. We disregard 9 (of 64) features that are not applicable in our setting.
Details about our re-implementation are given in Appendix~\ref{liu-et-al-reimplementation}.

\begin{table*}[t]
\small
\centering
\begin{tabular}{c@{\hskip 2mm}l@{\hskip 4mm}ccccccccc}
\hline
& & \multicolumn{4}{c}{\bf Cleaned Test Sample} & & \multicolumn{4}{c}{\bf Full Test Set} \\
\cline{3-6}
\cline{8-11}
& \bf Model & \bf P & \bf R & \bf F1 & \bf Acc & & \bf P & \bf R & \bf F1 & \bf Acc\\
\hline

\multirow{5}{*}{Baselines} & \HasOverlap{}(\Comment{}, \DeletedCode{}) & 77.7 & 72.0 & 74.7  & 75.7 & & 74.1 & 62.8 & 68.0 & 70.4 \\
& \citeauthor{Corazza18}~\shortcite{Corazza18} & 65.1 & 46.0 & 53.9 & 60.7 & & 63.7 & 47.8 & 54.6 & 60.3 \\
& \PosthocBert{} &  66.2 & 70.4 & 67.9 & 66.9 && 68.9 & 73.2 & 70.7 & 69.8 \\
& \JustInTimeBert{} & 65.5 & 80.9 & 72.3 & 69.0 && 67.4 & 76.8 & 71.6 & 69.6 \\
& \citeauthor{LiuOutdatedLine}~\shortcite{LiuOutdatedLine} & 77.6  & 74.0 & 75.8 & 76.3 && 77.5 & 63.8 & 70.0 & 72.6	\\
\hline

\multirow{3}{*}{Post hoc} &
\PosthocSeq{} &  58.9 & 68.0 & 63.0 & 60.3 && 60.6 & 73.4	& 66.3 & 62.8 \\
& \PosthocGraph{} & 60.6 & 70.2	& 65.0 & 62.2 && 62.6 & 72.6 & 67.2 & 64.6  \\
& \PosthocHybrid{} &  53.7 & 77.3 & 63.3 & 55.2 && 56.3 & 80.8 & 66.3 & 58.9 \\
\hline

\multirow{3}{*}{Just-In-Time} 
& \JustInTimeSeq{} & 83.8 & 79.3 & 81.5 & 82.0 & & 80.7 & 73.8 & 77.1 & 78.0\\
& \JustInTimeGraph{} &  84.7 & 78.4 & 81.4 & 82.0 & & 79.8 & 74.4 & 76.9 & 77.6\\
&  \JustInTimeHybrid{} &  87.1 & 79.6 & 83.1 & 83.8& & 80.9& 74.7 &77.7 &	78.5
 \\
\hline

\multirow{3}{*}{Just-In-Time + features} &
\JustInTimeSeq{} + features &  91.3 & 82.0 & 86.4 & 87.1 && 88.4 & 73.2 & 80.0 & \bf 81.8\\
& \JustInTimeGraph{} + features &  85.8 & \bf 87.1 & 86.4 & 86.3 && 83.8 & \bf 78.3 & \bf 80.9 & 81.5 \\
&  \JustInTimeHybrid{} + features & \bf 92.3 & 82.4 &  \bf 87.1 & \bf 87.8 & &  \bf 88.6 &  72.4 &  79.6 &  81.5 \\
\hline

\end{tabular}
\vspace{-5pt}
\caption{\label{table:main-table}Results for baselines, \Posthoc{}, and \JustInTime{} models. Differences in F1 and Acc between \JustInTime{} vs. baseline models, \JustInTime{} vs. \Posthoc{} models, and  \JustInTime{} + features vs. \JustInTime{} models are statistically significant.}
\end{table*}

\subsection{Our Models}
\label{subsection:models}
\noindent\textbf{Post hoc:} We consider three models, with different ways of encoding the method. \PosthocSeq{} encodes \NewCode{} with a GRU, \PosthocGraph{} encodes \NewTree{} with a GGNN, and \PosthocHybrid{} uses both. Multi-head attention in \PosthocHybrid{} is computed with the hidden states of the two encoders separately and then combined.

\noindent\textbf{Just-In-Time:} To allow fair comparison with the \Posthoc{} setting, these models are identical in structure to the models described above except that \EditCode{} is used instead of \NewCode{}. 

\noindent\textbf{Just-In-Time + features:}
Because injecting explicit knowledge can boost the performance of neural models~\cite{ChenExplicitFeatures,XuanExternalFeatures}, we investigate adding comment and code features to our approach. These are computed at the token/node-level and concatenated with embeddings before being passed to encoders.  Features are derived from prior work on comments and code~\cite{panthaplackel2020associating, panthaplackel2020update}, including linguistic (e.g., POS tags) and lexical (e.g., comment/code overlap) features.

\begin{table*}[t]
\centering
\small
\begin{tabular}{c@{\hskip 2mm}l@{\hskip 4mm}lllllllll}
\hline
& & \multicolumn{4}{c}{\bf Cleaned Test Sample} & & \multicolumn{4}{c}{\bf Full Test Set} \\
\cline{3-6}
\cline{8-11}
& \bf Model & \bf P & \bf R & \bf F1 & \bf Acc & & \bf P & \bf R & \bf F1 & \bf Acc\\
\hline

\multirow{3}{*}{\Return{}} &
\JustInTimeSeq{} + features & 88.5$\ssymbol{1}$ & 72.0$\ssymbol{1}$ & \bf 79.4$\ssymbol{1}$ & \bf 81.3$\ssymbol{1}$ && \bf 87.6$\ssymbol{1}$ & 73.3$\ssymbol{1}$ & 79.8$\ssymbol{1}$ & \bf 81.4$\ssymbol{1}$ \\
& \JustInTimeGraph{} + features & 81.2  & \bf 77.3 & 79.1$\ssymbol{1}$ & 79.7 && 82.2 & \bf 79.3 & \bf 80.6 & 80.9$\ssymbol{1}$ \\
&  \JustInTimeHybrid{} + features & \bf 88.7$\ssymbol{1}$ & 72.0$\ssymbol{1}$ & \bf 79.4$\ssymbol{1}$ & \bf 81.3$\ssymbol{1}$ && 87.3$\ssymbol{1}$ & 73.7$\ssymbol{1}$ & 79.8$\ssymbol{1}$  & \bf 81.4$\ssymbol{1}$ \\
\hline

\multirow{3}{*}{\Param{}} &
\JustInTimeSeq{} + features &  90.0 & \bf 95.3 & 92.5 & 92.3$\ssymbol{2}$ &&  92.2 & 88.3$\ssymbol{2}$ & 90.2 & 90.4 \\
& \JustInTimeGraph{} + features & \bf 96.5  & 92.0 & \bf 94.2 & \bf 94.3 && \bf 94.5 & \bf 89.0$\ssymbol{2}$ & \bf 91.7 & \bf 91.9 \\
&  \JustInTimeHybrid{} + features &  94.6 & 89.3 & 91.8 & 92.0$\ssymbol{2}$ && 93.3 & 85.9 & 89.4 & 89.9 \\
\hline

\multirow{3}{*}{Summary} &
\JustInTimeSeq{} + features & \bf 96.0  & 78.7 & 86.5$\ssymbol{4}$ & 87.7 &&  84.7$\ssymbol{4}$ & 58.3 & 69.0 & \bf 73.9$\ssymbol{4}$ \\
& \JustInTimeGraph{} + features & 80.8  & \bf 92.0 & 86.0$\ssymbol{4}$ & 85.0 && 76.0 & \bf 66.4 & \bf 70.6 & 72.5 \\
&  \JustInTimeHybrid{} + features & 93.7 & 86.0 & \bf 89.5 & \bf 90.0 && \bf 85.0$\ssymbol{4}$ & 57.0 & 68.1  & 73.5$\ssymbol{4}$ \\
\hline

\end{tabular}
\vspace{-5pt}
\caption{\label{table:comment-specific} Evaluating performance with respect to different types of comments. Scores are averaged across 3 random restarts, and scores for which the difference in performance is \textit{not} statistically significant are shown with identical symbols.}
\end{table*}

\subsection{Model Training}
Models are trained to minimize negative log likelihood. We use 2-layer BiGRU encoders (hidden dimension 64). GGNN encoders (hidden dimension 64) are rolled out for 8 message-passing steps, also use hidden dimension 64. We initialize comment and code embeddings, of dimension 64, with pretrained ones~\cite{panthaplackel2020update}. Edit embeddings are of dimension 8. Attention modules use 4 attention heads. We use a dropout rate of 0.6. Training ends if the validation F1 does not improve for 10 epochs.

\section{Intrinsic Evaluation}
We report common classification metrics: precision (P), recall (R), and F1 (w.r.t. the positive label) and accuracy (Acc), averaged across 3 random restarts. We also perform significance testing~\cite{berg-kirkpatrick-etal-2012-empirical}.

In Table~\ref{table:main-table}, we report results for baselines, \Posthoc{} and \JustInTime{} inconsistency detection models. In the \Posthoc{} setting, we find that our three models can achieve higher F1 scores than the bag-of-words approach proposed by ~\citeauthor{Corazza18}~\shortcite{Corazza18}; however, they underperform the \PosthocBert{} baseline and significantly underperform all \JustInTime{} models, including the simple rule-based baseline. This demonstrates the benefit of performing inconsistency detection in the \JustInTime{} setting, in which the code changes that trigger inconsistency are available. Additionally, by encoding the syntactic structures of the comment and code changes, our \JustInTime{} models outperform this rule-based baseline as well as all other baselines and \Posthoc{} approaches. While the \JustInTimeHybrid{} model achieves slightly higher scores (on the basis of F1 and accuracy) than \JustInTimeSeq{} and \JustInTimeGraph{}, the differences are not statistically significant.

Our \JustInTime{} models outperform the rule-based and feature-based baselines, without any hand-engineered rules or features. However, by incorporating surface features into our \JustInTime{} models, we can further boost performance (by statistically significant margins). This suggests that our approach can be used in conjunction with task-specific rules~\cite{icomment2007,aComment,tComment,ratol2017fragile} and feature sets~\cite{LiuOutdatedLine} to build improved systems for specific domains.

Furthermore, in Table~\ref{table:comment-specific},
we analyze the performance of the three \JustInTime{} + features models with respect to individual comment types. While these models are trained on all comment types together without explicitly tailoring it in any way to handle them differently, they are still able to achieve reasonable performance across types.
We provide further analysis of individual comment types and compare to comment-specific baselines in Appendix~\ref{appendix:comment-specific}.

\section{Extrinsic Evaluation}
We further evaluate how our approach could be used to build a comprehensive \emph{\JustInTime{} comment maintenance system} which first determines whether a comment, \Comment{}, has become inconsistent upon code changes to the corresponding method (\OldCode{} $\rightarrow$ \NewCode{}), and then automatically suggests an update if this is the case. To do this, we combine the inconsistency detection approach with our previously proposed comment update model~\cite{panthaplackel2020update} which updates comments based on code changes. For training and evaluating this combined system, we have two sets of comment/method pairs from consecutive commits for each example in our corpus. Recall from our data collection procedure that we extracted pairs of the form (\CommentA{}, \CodeA{}), (\CommentB{}, \CodeB{}), where \Comment{}=\CommentA{}, \OldCode{}=\CodeA{}, and \NewCode{}=\CodeB{}. We now introduce \NewComment{}=\CommentB{}, the gold comment for \NewCode{}. If \Comment{} is consistent with \NewCode{}, \Comment{}=\NewComment{}.

\subsection{Evaluation Method}
\label{sec:extrinsic-method}
The GRU-based \textsc{Seq2Seq} update model encodes \Comment{} and a sequential representation of the code changes (\EditCode{}). Using attention~\cite{Luong2015Attention} and a pointer network~\cite{VinyalsPointer} over learned representations of the inputs, a sequence of edit actions (\EditComment{}) is generated, identifying how \Comment{} should be edited to form the updated comment (\NewComment{}). This model also employs the same linguistic and lexical features as the \JustInTime{} + features models. The model is trained on only cases in which \Comment{} has to be updated and is not designed to ever copy the existing comment. We consider three different configurations for adding inconsistency detection in this model:

\noindent\textbf{\UpdateCopy{}:} We augment training of the update model with negative examples (i.e., \Comment{} does not need to be updated).
The model implicitly does inconsistency detection by learning to copy \Comment{} for such cases. Inconsistency detection is evaluated based on whether it predicts \NewComment{}=\Comment{}.

\noindent\textbf{\Pretrained{}:} The update model is \citet{panthaplackel2020update}, trained on only positive examples. At test time, if the detection model classifies \Comment{} as inconsistent, we take the prediction of the update model. Otherwise, we copy \Comment{}, making \NewComment{}=\Comment{}. We consider three of the pretrained \JustInTime{} detection models. 

\noindent\textbf{\JointlyTrained{}:} We jointly train the inconsistency detection and update models on the full dataset (including positive and negative examples). We consider three of our \JustInTime{} detection techniques. The update model and detection model share embeddings and the comment encoder for all three, and for the sequence-based and hybrid models, the code sequence encoder is also shared. During training, loss is computed as the sum of the update and detection components. For negative examples, we mask the loss of the update component since it does not have to learn to copy \Comment{}. At test time, if the detection component predicts a negative label, we directly copy \Comment{} and otherwise take the prediction of the update model.

\begin{table*}
\small
\centering
\begin{tabular}{l@{\hskip 3mm}llllll@{\hskip 1mm}llll}
\hline
& \multicolumn{5}{c}{\bf Update Metrics} & & \multicolumn{4}{c}{\bf Detection Metrics} \\
\cline{2-6}
\cline{8-11}
& \bf xMatch & \bf METEOR & \bf BLEU-4 & \bf SARI & \bf GLEU & & \bf P & \bf R & \bf F1 & \bf Acc \\
\hline
Never Update & 50.0 & 67.4 & 72.1 & 24.9 & 68.2 & & 0.0 & 0.0 & 0.0 & 50.0 \\
\citeauthor{panthaplackel2020update}~\shortcite{panthaplackel2020update} &  25.9 & 60.0 & 68.7 & 42.0$\ssymbol{1}$ &	67.4 && 54.0 & \bf 95.6 & 69.0 & 57.1 \\
\hline

\UpdateCopy{} & 58.0 & 72.0 & 74.7 & 31.5 & 72.7 && \bf 100.0 & 23.3 &	37.7 & 61.7 \\

\hline

\Pretrained{} &  &  &  &  &  & &  &  &   &  \\
\hspace{0.2cm}\JustInTimeSeq{} + features &  \bf 62.3$\ssymbol{2}$ & 75.6$\ssymbol{1}$ & 77.0$\ssymbol{1}$ & 42.0$\ssymbol{1}$ & 76.2 && 91.3$\ssymbol{1}$ & 82.0$\ssymbol{4}$ & 86.4$\ssymbol{1}$ & 87.1$\ssymbol{4}$$\ssymbol{5}$  \\
\hspace{0.2cm}\JustInTimeGraph{} + features &  59.4 &  74.9$\ssymbol{4}$ & 76.6$\ssymbol{2}$ & \bf 42.5$\ssymbol{6}$ & 75.8$\ssymbol{1}$$\ssymbol{2}$ && 85.8 & 87.1 & 86.4$\ssymbol{1}$ & 86.3$\ssymbol{2}$
\\
\hspace{0.2cm}\JustInTimeHybrid{} + features & \bf 62.3$\ssymbol{2}$ & 75.8$\ssymbol{2}$$\ssymbol{6}$ & \bf 77.2 & 42.3$\ssymbol{2}$ & \bf 76.4 && 92.3 & 82.4$\ssymbol{4}$ & 87.1$\ssymbol{2}$ & 87.8$\ssymbol{1}$$\ssymbol{6}$	\\
\hline

\JointlyTrained{} &  &  &  &  &  & &  &  &   &  \\
\hspace{0.2cm}\JustInTimeSeq{} + features &  61.4$\ssymbol{1}$ & \bf 75.9$\ssymbol{6}$ & 76.6$\ssymbol{2}$ & 42.4$\ssymbol{2}$$\ssymbol{6}$ & 75.6$\ssymbol{2}$ && 88.3$\ssymbol{2}$ & 86.2 & 87.2$\ssymbol{2}$ & 87.3$\ssymbol{4}$$\ssymbol{6}$ \\
\hspace{0.2cm}\JustInTimeGraph{} + features &  60.8 & 75.1$\ssymbol{4}$ & 76.6$\ssymbol{2}$  & 41.8$\ssymbol{1}$ & 75.8$\ssymbol{1}$ && 88.3$\ssymbol{2}$ & 84.7$\ssymbol{1}$ & 86.4$\ssymbol{1}$ & 86.7$\ssymbol{2}$$\ssymbol{5}$ \\
\hspace{0.2cm}\JustInTimeHybrid{} + features & 61.6$\ssymbol{1}$ & 75.6$\ssymbol{1}$$\ssymbol{2}$ & 76.9$\ssymbol{1}$ & 42.3$\ssymbol{2}$ & 75.9$\ssymbol{1}$ && 90.9$\ssymbol{1}$ & 84.9$\ssymbol{1}$ & \bf 87.8 & \bf 88.2$\ssymbol{1}$ \\
\hline
\end{tabular}
\vspace{-5pt}
\caption{\label{table:comment-update-table}Results on joint inconsistency detection and update on the cleaned test sample. Scores for which the difference in performance is \textit{not} statistically significant are shown with identical symbols.}
\end{table*}

\subsection{Results}
\label{sec:extrinsic-results}
We report precision, recall, F1, and accuracy for detection. As we have done previously~\cite{panthaplackel2020update}, we evaluate update through exact match (xMatch) as well as metrics used to evaluate text generation (BLEU-4~\cite{papineni2002bleu} and METEOR~\cite{BanerjeeEtAL2005}) and text editing tasks (SARI~\cite{xu-etal-2016-optimizing} and GLEU~\cite{napoles-etal-2015-ground}). In Table~\ref{table:comment-update-table}, we compare performances of combined inconsistency detection and update systems on the cleaned test sample. As reference points, we also provide scores for a system which never updates (i.e., always copies \Comment{} as \NewComment{}) and  ~\citeauthor{panthaplackel2020update}~\shortcite{panthaplackel2020update}, which is designed to always update (and only copy \Comment{} if an invalid edit action sequence is generated).
For completeness, we also provide results on the full dataset (which are analogous) in Appendix~\ref{appendix:combined}. 

Since our dataset is balanced, we can get 50\% exact match by simply copying \Comment{} (i.e., never updating). In fact, this can even beat \citeauthor{panthaplackel2020update}~\shortcite{panthaplackel2020update} on xMatch, METEOR, BLEU-4, SARI, and GLEU. This underlines the importance of first determining whether a comment needs to be updated, which can be addressed with our inconsistency detection approach. On the  majority of the update metrics, both of these underperform the other three approaches (\UpdateCopy{}, \Pretrained{}, and \JointlyTrained{}). SARI is calculated by averaging N-gram F1 scores for edit operations (add, delete, and keep). So, it is not surprising that the \textit{\UpdateCopy{}} baseline, which learns to copy, performs fewer edits, consequently underperforming on this metric. Because \citeauthor{panthaplackel2020update}~\shortcite{panthaplackel2020update} is designed to \textit{always} edit, it can perform well on this metric; however, the majority of the pretrained and jointly trained systems can beat this.

The \textit{\UpdateCopy{}} baseline, which does not include an explicit inconsistency detection component, performs relatively well with respect to the update metrics, but it performs poorly on detection metrics. Here, we use generating \Comment{} as the prediction for \NewComment{} as a proxy for detecting inconsistency. It achieves high precision, but it frequently copies \Comment{} in cases in which it is inconsistent and should be updated, hence underperforming on recall. The pretrained and jointly trained approaches outperform this model by wide statistically significant margins across the majority of metrics, demonstrating the need for inconsistency detection. 

We do not observe a significant difference between the pretrained and jointly trained systems. The pretrained models achieve slightly higher scores on most update metrics and the jointly trained models achieve slightly higher scores on the detection metrics; however, these differences are small and often statistically insignificant. Overall, we find that our approach can be useful for building a real-time comment maintenance system. Since this is not the focus of our paper but rather merely a potential use case,  we leave it to future work for developing more intricate joint systems.

\section{Related Work}
\noindent\textbf{Code/Comment Inconsistencies:}
Prior work analyze how inconsistencies emerge ~\cite{FluriAnalysis,JiangEvolution,IbrahimBugs,fluri:coevolve} and the various types of inconsistencies~\cite{WenLargeStudy}; but, they do not propose techniques for addressing the problem.

\noindent\textbf{Post Hoc Inconsistency Detection:}
Prior work propose rule-based approaches for detecting pre-existing inconsistencies in specific domains, including locks~\cite{icomment2007}, interrupts~\cite{aComment}, \CodeIn{null} exceptions for method parameters~\cite{ZhouParameter,tComment}, and renamed identifiers~\cite{ratol2017fragile}. The comments they consider are consequently constrained to certain templates relevant to their respective domains. We instead develop a general-purpose, machine learning approach that is not catered towards any specific types of inconsistencies or comments. \citeauthor{Corazza18}~\shortcite{Corazza18} and~\citet{Cimasa19} address a broader notion of coherence between comments and code through text-similarity techniques, and~\citet{Khamis2010AutomaticQA} determine whether comments, specifically \Return{} and \Param{} comments, conform to particular format. We instead capture deeper code/comment relationships by learning their syntactic and semantic structures. ~\citet{RabbiInconsistencySiamese} propose a siamese network
for correlating comment/code representations. In contrast, we aim to correlate comments and code through an attention mechanism.

\noindent\textbf{Just-In-Time Inconsistency Detection:}
~\citeauthor{LiuOutdatedLine}~\shortcite{LiuOutdatedLine} detect inconsistencies in a block/line comment upon changes to the corresponding code snippet using a random forest classifier with hand-engineered features. Our approach does not require such extensive feature engineering. Although their task is slightly different, we consider their approach as a baseline.
\citet{StulovaTowards} concurrently present a preliminary study of an approach which maps a comment to the AST nodes of the method signature (before the code change) using BOW-based similarity metrics. This mapping is used to determine whether the code changes have triggered a comment inconsistency. Our model instead leverages the full method context and also learns to map the comment directly to the code changes. \citet{Malik08} predict whether a comment will be updated using a random forest classifier utilizing surface features that capture aspects of the method that is changed, the change itself, and ownership. They do not consider the existing comment since their focus is not inconsistency detection; instead, they aim to understand the rationale behind comment updating practices by analyzing useful features. ~\citet{SaduThesis} develops at approach which locates inconsistent identifiers upon code changes through lexical matching rules. While we find such a rule-based approach (represented by our \HasOverlap{}(\Comment{}, \DeletedCode{}) baseline) to be effective, a learned model performs significantly better. ~\citet{Svensson2015ReducingOA} builds a system to mitigate the damage of inconsistent comments by prompting developers to validate a comment upon code changes. Comments that are not validated are identified, indicating that they may be out of date and unreliable. \citet{NieTrigit} present a framework for maintaining consistency between code and todo comments by performing actions described in such comments when code changes trigger the specified conditions to be satisfied.

\section{Conclusion}
We developed a deep learning approach for \JustInTime{} inconsistency detection between code and comments by learning to relate comments and code changes. Based on evaluation on a large corpus consisting of multiple types of comments, we showed that our model substantially  outperforms various baselines as well as \Posthoc{} models that do not consider code changes. We further conducted an extrinsic evaluation in which we demonstrated that our approach can be used to build a comprehensive comment maintenance system that can detect and update inconsistent comments.

\section*{Acknowledgments}
This work was supported by the Bloomberg Data Science Fellowship and a Google Faculty Research Award.

\section*{Ethics Statement}
Through this work, we aim to reduce time-consuming confusion and vulnerability to software bugs by keeping developers informed with up-to-date-documentation, in order to consequently help improve developers’ productivity and software quality. Buggy software and incorrect API usage can result in significant malfunctions in many everyday operations. Maintaining comment/code consistency can help prevent such negative-impact events.

However, over-reliance on such a system could result in developers giving up identifying and resolving inconsistent comments themselves. By presuming that the system detects all inconsistencies and all of these are properly addressed, developers may also take the available comments for granted, without carefully analyzing their validity. Because the system may not catch all types of inconsistencies, this could potentially exacerbate rather than resolve the problem of inconsistent comments. Our system is not intended to serve as an infallible safety net for poor software engineering practices but rather as a tool that complements good ones, working alongside developers to help deliver reliable, well-documented software in a timely manner.

\begin{small}
\bibliography{references}
\end{small}
\clearpage
\newpage
\appendix

\section{Additional Data Details}
\label{appendix:data}
We provide additional information about our procedures for filtering the dataset and curating a sample of the test set for evaluation. We also include various statistics about the examples that comprise our dataset.

\subsection{Filtering}
Recall from our data collection procedure (\S\ref{sec:data}) that we extracted comment/method pairs from consecutive commits, of the form (\CommentA{}, \CodeA{}), (\CommentB{}, \CodeB{}), where \Comment{}=\CommentA{}, \OldCode{}=\CodeA{}, and \NewCode{}=\CodeB{}. We apply heuristics to reduce the number of cases in which there are unrelated comment and code changes. We filter out positive examples in which the differences between \CommentA{} and \CommentB{} entail minor cosmetic edits (e.g., reformatting, spelling corrections). Similar to prior work~\cite{panthaplackel2020update}, for \Return{} examples, we require there to be a code change to at least one return statement or the return type of the method. We discard all \Return{} examples (positive and negative) that do not satisfy this condition. This is because \Return{} comments describe aspects of the return value of a method, which is typically related to the method return type and return statements. We apply the same constraint to
summary comment examples, since they often describe aspects of the output (e.g., Figure~\ref{fig:apache-ignite}). Because \Param{} comments generally pertain to the method's arguments, we only use examples in which an argument name or type is changed within the method. Next, we reduce the number of noisy negative examples in which a developer fails to update comments in accordance with code changes by limiting to the top 1,000 starred and forked projects, which are considered popular and well-maintained~\cite{ProjectQuality}. Furthermore, because we consider AST representations, we remove all examples consisting of a method which cannot be parsed into an AST. Additionally, we remove duplicate examples, as they have been found to negatively affect training machine learning models for source code~\cite{allamanis2019duplication}.

\subsection{Downsampling Negative Class}

\begin{table}
\centering
\small
\begin{tabular}{l@{\hskip 3mm}lll}
\hline
& \bf Positive & \bf Negative & \bf Total \\
\hline
\Return{} & 9,807 & 72,826 & 82,633 \\
\Param{} & 5,507 & 19,007 & 24,514 \\
Summary & 5,904 & 69,650 & 75,554 \\
\hline
Full & 21,218 & 161,483 & 182,701\\
\hline
\end{tabular}
\vspace{-5pt}
\caption{\label{table:downsample}Dataset sizes before downsampling}
\vspace{-10pt}
\end{table}

In our data collection procedure, we obtained substantially more negative examples.
Because a na\"ive classifier trained to always predict the negative label can achieve high accuracy in such a setting, we downsample the negative class to obtain a balanced dataset. We provide the sizes of the positive and negative classes before downsampling in Table~\ref{table:downsample}. Note that we also discard some examples to ensure no overlap between the projects in training, validation, and test.

\subsection{Curating Test Sample}
\label{appendix:annotation}
We construct a clean test set by randomly sampling without replacement from the full test set in such a way that we attain a balanced sample, in terms of both labels (i.e., positive, negative) and comment type (i.e., \Return{}, \Param{}, summary). Rather than assigning new labels to mislabeled examples, we choose to remove such examples altogether from this sample. Note that for examples that are incorrectly labeled as negative and should actually be updated, we do not have ground truth updated comments (i.e., \NewComment{}) for evaluation. So, by removing mislabeled examples, we can use the same set of examples to evaluate the combined inconsistency detection and update system. %
This cleaning procedure was done by one of the authors of this paper who has 8+ years of Java experience.

We remove 11\% of examples for having the incorrect label, 3\% for being uncertain about the correct label due to the limited context provided in the method, and 6\% from being poor examples (e.g., comments like ``document me'' or code changes that simply comment out the entire method). Therefore, we find ~17-20\% noise. For the individual comment types, the percent of noise is 6-15\% for \Return{}, 14-16\% for \Param{}, and 26-28\% for summary comments. For individual labels, it is 26-28\% for positive and 8-12\% for negative.

\subsection{Data Statistics}

\begin{table}
\centering
\small
\begin{tabular}{l@{\hskip 3mm}rrrrr}
\hline
& \Return{} & \Param{} & Summary & Full \\ 
\hline
\bf \Comment{} & 9.7 & 8.4 & 13.3 & 10.3\\
\bf \OldCode{} & 131.1 & 186.9 & 137.0 & 147.2\\
\bf \NewCode{} & 131.9 & 187.7 & 135.4 & 147.3\\
\bf \EditCode{} & 179.4 & 240.9 & 186.6 & 197.3\\
\bf \OldTree{} & 127.2 & 184.1 & 130.5 & 142.9 \\
\bf \NewTree{} & 128.1 & 184.5 & 129.5 & 143.2 \\
\bf \EditTree{} & 154.3 & 213.7 & 159.1 & 171.1 \\

\hline
\end{tabular}
\vspace{-5pt}
\caption{\label{table:data-stats} Statistics on the average lengths of comment and code representations.}
\end{table}

In Table~\ref{table:data-stats}, we show the average lengths of comment and code representations for the various types of comments in our dataset. The lengths for \Comment{} and sequential code representations (i.e., \OldCode{}, \NewCode{}, \EditCode{}) are computed based on the subtokenized sequences that are used by our model. Note that the \EditCode{} representation also includes edit keywords. We report the sizes of the AST representations (\OldTree{}, \NewTree{}, \EditTree{}) in terms of number of nodes. This also includes the added \textit{subnodes}.

\begin{table*}[ht]
\centering
\small
\begin{tabular}{ll@{\hskip 3mm}lllllllll}
\hline
&& \multicolumn{4}{c}{\bf Cleaned Test Sample} & & \multicolumn{4}{c}{\bf Full Test} \\
\cline{3-6}
\cline{8-11}
& \bf Model & \bf P & \bf R & \bf F1 & \bf Acc & & \bf P & \bf R & \bf F1 & \bf Acc\\
\hline
\multirow{10}{*}{\Return{}} & 
\HasOverlap{}(\Comment{}, \DeletedCode{})  & 69.2 & 54.0 & 60.7 & 65.0$\ssymbol{6}$ & & 67.6$\ssymbol{6}$$\ssymbol{5}$ & 53.3$\ssymbol{6}$ & 59.6 & 63.9 \\
& \citeauthor{Corazza18}~\shortcite{Corazza18}  & 73.2 & 60.0$\ssymbol{5}$ & 65.9 & 69.0 & & 68.9$\ssymbol{5}$ & 61.2$\ssymbol{4}$ & 64.8 & 66.8 \\
& \PosthocBert{}  & 84.9$\ssymbol{1}$ & 74.7$\ssymbol{4}$ & 79.4$\ssymbol{1}$$\ssymbol{5}$ & 80.7$\ssymbol{1}$ & & 85.6 & 82.7 & \bf 84.1 & \bf 84.3 \\
& \JustInTimeBert{}  & 62.5 & 74.0$\ssymbol{4}$& 67.7$\ssymbol{6}$ & 64.7$\ssymbol{6}$ & & 66.8$\ssymbol{6}$ & 78.8$\ssymbol{1}$$\ssymbol{2}$ & 72.2 & 69.7 \\
& \citeauthor{LiuOutdatedLine}~\shortcite{LiuOutdatedLine}  & 76.9 & 62.0$\ssymbol{5}$ & 68.6$\ssymbol{6}$ & 71.7 & & 76.0 & 63.0$\ssymbol{4}$ & 68.9 & 71.6 \\
& \citet{Khamis2010AutomaticQA}  & 52.1 & \bf 98.0 & 68.1$\ssymbol{6}$ & 54.0 & & 51.6 & \bf 97.3 & 67.4 & 52.9 \\
& \GenMatch{}  & 64.6 & 62.0$\ssymbol{5}$ & 63.3 & 64.0$\ssymbol{6}$ & & 60.4 & 54.9$\ssymbol{6}$ & 57.5 & 59.5 \\
\cline{2-11}
& \JustInTimeSeq{} + features  & 85.3$\ssymbol{1}$ & 75.3$\ssymbol{4}$ & 79.9$\ssymbol{5}$ & 81.0$\ssymbol{1}$ & & 87.2$\ssymbol{4}$ & 75.9 & 81.2$\ssymbol{1}$$\ssymbol{4}$ & 82.4 \\
& \JustInTimeGraph{} + features  & 87.4 & 77.3$\ssymbol{1}$ & \bf 82.0 & \bf 83.0 & & 84.0$\ssymbol{1}$ & 78.0$\ssymbol{1}$ & 80.8$\ssymbol{1}$$\ssymbol{4}$ & 81.4$\ssymbol{1}$$\ssymbol{2}$ \\
& \JustInTimeHybrid{} + features  & 84.8$\ssymbol{1}$ & 78.0$\ssymbol{2}$ &  81.2 & 82.0$\ssymbol{5}$ & & 84.3$\ssymbol{1}$ & 78.7$\ssymbol{2}$& 81.3$\ssymbol{4}$ & 81.9$\ssymbol{1}$\\

\hline
\multirow{3}{*}{Combined} &
\JustInTimeSeq{} + features  & 88.5$\ssymbol{4}$ & 72.0$\ssymbol{6}$ & 79.4$\ssymbol{1}$$\ssymbol{5}$ & 81.3$\ssymbol{1}$$\ssymbol{5}$ & & \bf 87.6$\ssymbol{4}$ & 73.3$\ssymbol{5}$ & 79.8$\ssymbol{5}$ & 81.4$\ssymbol{1}$$\ssymbol{2}$ \\
& \JustInTimeGraph{} + features  & 81.2 & 77.3$\ssymbol{1}$$\ssymbol{2}$ & 79.1$\ssymbol{1}$ & 79.7 & & 82.2 & 79.3$\ssymbol{2}$ & 80.6$\ssymbol{1}$ & 80.9$\ssymbol{2}$ \\
& \JustInTimeHybrid{} + features  & \bf 88.7$\ssymbol{4}$ & 72.0$\ssymbol{6}$ & 79.4$\ssymbol{1}$ & 81.3$\ssymbol{1}$ & & 87.3$\ssymbol{4}$ & 73.7$\ssymbol{5}$ & 79.8$\ssymbol{5}$ & 81.4$\ssymbol{1}$$\ssymbol{2}$ \\

\hline

\end{tabular}
\vspace{-5pt}
\caption{\label{table:return-table} Results for \Return{} examples. Scores for which the difference in performance is \textit{not} statistically significant are shown with identical symbols.}
\end{table*}

\begin{table*}[ht]
\centering
\small
\begin{tabular}{ll@{\hskip 3mm}lllllllll}
\hline
&& \multicolumn{4}{c}{\bf Cleaned Test Sample} & & \multicolumn{4}{c}{\bf Full Test} \\
\cline{3-6}
\cline{8-11}
& \bf Model & \bf P & \bf R & \bf F1 & \bf Acc & & \bf P & \bf R & \bf F1 & \bf Acc\\
\hline
\multirow{9}{*}{\Param{}} 
& \HasOverlap{}(\Comment{}, \DeletedCode{})  & 85.7 & \bf 96.0$\ssymbol{1}$$\ssymbol{4}$ & 90.6 & 90.0 & & 84.0 & \bf 93.3 & 88.4$\ssymbol{5}$ & 87.8$\ssymbol{5}$ \\
& \citeauthor{Corazza18}~\shortcite{Corazza18}  & 74.1 & 40.0 & 51.9 & 63.0 & & 59.1$\ssymbol{4}$ & 43.9 & 50.4 & 56.7 \\
& \PosthocBert{}  & 62.8 & 57.3 & 59.9 & 61.7 & & 58.9$\ssymbol{4}$ & 64.4 & 61.5 & 59.7 \\
& \JustInTimeBert{}  & 81.8 & 84.0 & 82.8 & 82.7 & & 75.5 & 82.7 & 78.9$\ssymbol{4}$ & 77.9 \\
& \citeauthor{LiuOutdatedLine}~\shortcite{LiuOutdatedLine}  & 90.4$\ssymbol{4}$ & 62.7 & 74.0 & 78.0 & & 88.6$\ssymbol{5}$ & 72.3 & 79.6$\ssymbol{4}$ & 81.5 \\

& \citet{Khamis2010AutomaticQA}  & \bf 97.8$\ssymbol{1}$ & 90.0$\ssymbol{6}$ & 93.8 & 94.0$\ssymbol{2}$ & & 87.7$\ssymbol{5}$ & 89.0$\ssymbol{1}$$\ssymbol{4}$ & 88.3$\ssymbol{5}$ & 88.2$\ssymbol{5}$ \\
\cline{2-11}
& \JustInTimeSeq{} + features  & 95.4 & \bf 96.0$\ssymbol{1}$ & 95.7$\ssymbol{1}$$\ssymbol{2}$ & 95.7$\ssymbol{1}$ & & 91.4 & 89.2$\ssymbol{4}$ & 90.3$\ssymbol{2}$ & 90.4$\ssymbol{2}$$\ssymbol{4}$ \\

& \JustInTimeGraph{} + features  & 97.3$\ssymbol{1}$ & 94.0 & 95.6$\ssymbol{1}$ & 95.7$\ssymbol{1}$ & & \bf 94.9$\ssymbol{1}$ & 90.0 & \bf 92.4 & \bf 92.6 \\

& \JustInTimeHybrid{} + features  & 96.6$\ssymbol{2}$ & 95.3$\ssymbol{4}$ & \bf 96.0$\ssymbol{2}$ & \bf 96.0 & & 94.3$\ssymbol{2}$ & 89.3$\ssymbol{4}$ & 91.7$\ssymbol{1}$ & 91.9$\ssymbol{1}$ \\

\hline
\multirow{3}{*}{Combined}
& \JustInTimeSeq{} + features  & 90.0$\ssymbol{4}$ & 95.3$\ssymbol{4}$ & 92.5 & 92.3$\ssymbol{4}$ & & 92.2 & 88.3$\ssymbol{1}$ & 90.2$\ssymbol{2}$ & 90.4$\ssymbol{2}$ \\
&  \JustInTimeGraph{} + features  & 96.5$\ssymbol{2}$ & 92.0 & 94.2 & 94.3$\ssymbol{2}$ & & 94.5$\ssymbol{1}$$\ssymbol{2}$ & 89.0$\ssymbol{1}$$\ssymbol{4}$ & 91.7$\ssymbol{1}$ & 91.9$\ssymbol{1}$ \\
& \JustInTimeHybrid{} + features  & 94.6 & 89.3$\ssymbol{6}$ & 91.8 & 92.0$\ssymbol{4}$ & & 93.3 & 85.9 & 89.4 & 89.9$\ssymbol{4}$ \\

\hline

\end{tabular}
\vspace{-5pt}
\caption{\label{table:param-table}Results for \Param{} examples. Scores for which the difference in performance is \textit{not} statistically significant are shown with identical symbols.}
\end{table*}

\section{Re-Implementation of ~\citet{LiuOutdatedLine}}
\label{liu-et-al-reimplementation}
From the 64 features used in ~\citet{LiuOutdatedLine}, we disregard 9 which are not compatible with our setting. Namely, since the body of code they consider are not full methods, they have separate features for the code snippet under consideration and the method to which it belongs. On the other hand, we look at only full methods, and so the separate notion of ``code snippet" is irrelevant. They have two separate features \textit{comment: ratio of comment lines to the method, comment: ratio of comment lines to the code snippet}, and we discard the latter. Next, they consider whether there is a name change for the method in which the code snippet appears; however, we look at two versions of the same method and require the name to remain the same (as part of our data collection procedure). So, we discard the feature \textit{refactoring: rename method}. Additionally, we discard external features that are not extracted from outside the method (e.g., class-related features) as we focus on detecting inconsistencies using only the method-level context. We leave it to future work to study ways to incorporate the external context into our approach. We discard \textit{code: changes on class attribute, code: class attribute related, refactoring: extract method, refactoring: inline method, refactoring: encapsulate field, refactoring: replace exception with test, comment: ratio of comment lines to class}.

\begin{table*}[ht]
\centering
\small
\begin{tabular}{ll@{\hskip 3mm}lllllllll}
\hline
&& \multicolumn{4}{c}{\bf Cleaned Test Sample} & & \multicolumn{4}{c}{\bf Full Test} \\
\cline{3-6}
\cline{8-11}
& \bf Model & \bf P & \bf R & \bf F1 & \bf Acc & & \bf P & \bf R & \bf F1 & \bf Acc\\
\hline
\multirow{8}{*}{Summary} 
& \HasOverlap{}(\Comment{}, \DeletedCode{})  & 75.0$\ssymbol{1}$ & 66.0 & 70.2$\ssymbol{4}$ & 72.0 & & 71.4 & 49.7 & 58.6 & 64.9$\ssymbol{4}$ \\
& \citeauthor{Corazza18}~\shortcite{Corazza18}  & 55.6 & 30.0 & 39.0 & 53.0 & & 61.7 & 41.1 & 49.3 & 57.8 \\
& \PosthocBert{}  & 61.6$\ssymbol{2}$ & 78.7$\ssymbol{2}$ & 68.8 & 64.3 & & 63.7$\ssymbol{4}$ & \bf 75.6$\ssymbol{1}$ & 68.9$\ssymbol{1}$$\ssymbol{2}$$\ssymbol{4}$ & 66.0$\ssymbol{4}$$\ssymbol{2}$ \\
& \JustInTimeBert{}  & 62.1$\ssymbol{2}$ & 80.0$\ssymbol{4}$ & 69.8$\ssymbol{4}$ & 65.3 & & 64.5$\ssymbol{4}$ & 72.4$\ssymbol{4}$ & 68.0$\ssymbol{2}$$\ssymbol{4}$ & 65.9$\ssymbol{4}$$\ssymbol{2}$ \\
& \citeauthor{LiuOutdatedLine}~\shortcite{LiuOutdatedLine}  & 85.2 & 76.7 & 80.7$\ssymbol{2}$ & 81.7 & & 77.1$\ssymbol{1}$ & 57.0$\ssymbol{2}$$\ssymbol{5}$ & 65.5 & 70.0 \\

\cline{2-11}
& \JustInTimeSeq{} + features  & 72.7 & \bf 92.7 & 81.4$\ssymbol{2}$ & 78.7 & & 67.7 & 74.3$\ssymbol{1}$ & \bf 70.6$\ssymbol{6}$ & 68.9 \\
& \JustInTimeGraph{} + features  & 74.3$\ssymbol{1}$ & 92.0$\ssymbol{1}$ & 82.0 & 79.3 & & 68.4 & 70.9 & 69.2$\ssymbol{1}$ & 68.2 \\
& \JustInTimeHybrid{} + features  & 70.7 & 90.0 & 79.2 & 76.3 & & 64.5$\ssymbol{4}$ & 72.9$\ssymbol{4}$ & 68.4$\ssymbol{2}$$\ssymbol{4}$ & 66.3$\ssymbol{2}$ \\

\hline
\multirow{3}{*}{Combined}
& \JustInTimeSeq{} + features  & \bf 96.0 & 78.7$\ssymbol{2}$$\ssymbol{4}$ & 86.5$\ssymbol{1}$ & 87.7 & & 84.7$\ssymbol{2}$ & 58.3$\ssymbol{2}$ & 69.0$\ssymbol{1}$$\ssymbol{2}$ & \bf 73.9$\ssymbol{1}$ \\
& \JustInTimeGraph{} + features  & 80.8 & 92.0$\ssymbol{1}$ & 86.0$\ssymbol{1}$ & 85.0 & & 76.0$\ssymbol{1}$ & 66.4 & \bf 70.6$\ssymbol{6}$ & 72.5 \\
& \JustInTimeHybrid{} + features  & 93.7 & 86.0 & \bf 89.5 & \bf 90.0 & & \bf 85.0$\ssymbol{2}$ & 57.0$\ssymbol{5}$ & 68.1$\ssymbol{4}$ & 73.5$\ssymbol{1}$ \\

\hline

\end{tabular}
\vspace{-5pt}
\caption{\label{table:summary-table} Results for summary comment examples. Scores for which the difference in performance is \textit{not} statistically significant are shown with identical symbols.}
\end{table*}

\section{Comment-Specific Performance}
\label{appendix:comment-specific}
Since much of the work in inconsistency detection has focused on comment-specific or task-specific settings, we analyze how our approach performs in a similar setting. Namely, we consider training and evaluating our models and baselines on only data pertaining to individual comment types. We additionally study how comment-specific training compares to combined training, in which we train on the full dataset, comprised of multiple comment types.

\subsection{\Return{} Comments}
We train the (learned) baselines introduced in Section~\ref{sec:baselines} on only the 15,950 examples pertaining to \Return{} comments. We additionally consider two baselines for \Return{} comments. \citet{Khamis2010AutomaticQA} proposed a heuristic for detecting inconsistency in \Return{} comments: the comment must begin with the correct return type of the corresponding method. We implement a baseline based on this heuristic. We also remove articles (e.g., \emph{a}, \emph{the}) from the beginning of the comment before applying this rule, as we found this to improve performance. Furthermore,~\citeauthor{panthaplackel2020update}~\shortcite{panthaplackel2020update} released an \Return{} comment generation model trained on a large corpus of \Return{} comment/method pairs. We introduce another baseline, \GenMatch{}, in which we use the pretrained generation model to generate an \Return{} comment for \OldCode{} and an \Return{} comment for \NewCode{}. If the two comments match exactly,
we consider the code change to be irrelevant to \Return{} comments and thus the existing \Return{} comment remains consistent. We compare these baselines with our models, trained on only \Return{} comments. We additionally compare with our models, trained on the combined training set, as done in the main paper.

In Table~\ref{table:return-table}, we report results on the 100 \Return{} examples in the cleaned test set as well as the 1,840 \Return{} examples in the full test set. While the \JustInTimeBert{} baseline performs quite well here, our approach can outperform baselines (w.r.t. F1 and Acc) on the cleaned test sample, when trained on only \Return{} comments. We find that training on the combined dataset slightly deteriorates performance of our models. This is not surprising as in combined training, models must learn to generalize across comment types, not just \Return{} comments. Nonetheless, the difference in performance between training on the comment-specific and combined sets are relatively small.

\subsection{\Param{} Comments}
For \Param{} comments, we consider another baseline designed to follow the heuristic proposed by~\citet{Khamis2010AutomaticQA} for this comment type: the comment should begin with the name of the parameter being documented. We remove articles from the beginning of the comment and consider whether the first term is one of the arguments of the method. If this is not the case, we classify it as inconsistent. We consider the comment-specific and combined settings, as we do for \Return{} comments.

In Table~\ref{table:param-table}, we report results on the 100 \Param{} examples in the cleaned test set as well as the 1,038 \Param{} examples in the full test set. We find that rule-based baselines can perform very well for \Param{} comments, especially the~\citet{Khamis2010AutomaticQA} baseline. This suggests that most \Param{} comments conform to the format they suggested. Nonetheless, our models are able to \textit{learn} this without explicitly specifying this format and can even achieve higher performance (by statistically significant margins) when trained on only \Param{} comments. The combined setting slightly deteriorates performance of our models; however, the \JustInTimeGraph{} + features model can still perform slightly better than ~\citet{Khamis2010AutomaticQA} w.r.t. F1 on the cleaned test sample.

 \begin{figure}[t]
\centering
\includegraphics[scale=0.25]{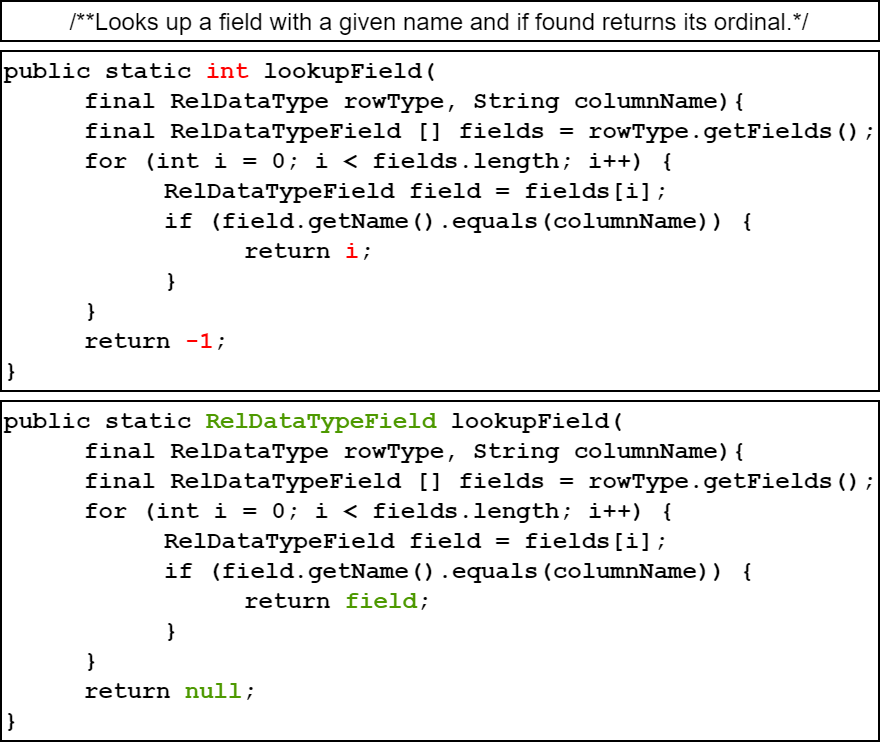}
\vspace{-5pt}
\caption{A summary comment from the Apache Calcite Avatica project becomes inconsistent upon changes to the \CodeIn{lookupField()} method. It should be updated to be \textit{Looks up a field with a given name, returning null if not found.}}
\label{fig:apache-calcite-avatica}
\vspace{-10pt}
\end{figure}

\subsection{Summary Comments}
Summary comments (e.g., Figure~\ref{fig:apache-calcite-avatica}) do not have a well-defined structure, and thus we do not have a format-based baseline as we did for \Return{} and \Param{} comments. We evaluate baselines and our models, trained on comment-specific data, as well as our model trained on the combined training set.

In Table~\ref{table:summary-table}, we report results on the 100 summary examples in the cleaned test set as well as the 1,066 summary examples in the full test set. While ~\citet{LiuOutdatedLine} is a strong baseline here, we find that we can outperform all baselines in the combined training setting. Unlike the case for \Return{} and \Param{} comments, combined training appears to yield improved performance over comment-specific training for our models. This suggests that the models can extract valuable information from the more structured comments in the training set that pertain to specific parts of the code in order to address the less-structured summary comments. 

\begin{figure*}[t!]
\centering
\includegraphics[scale=0.23]{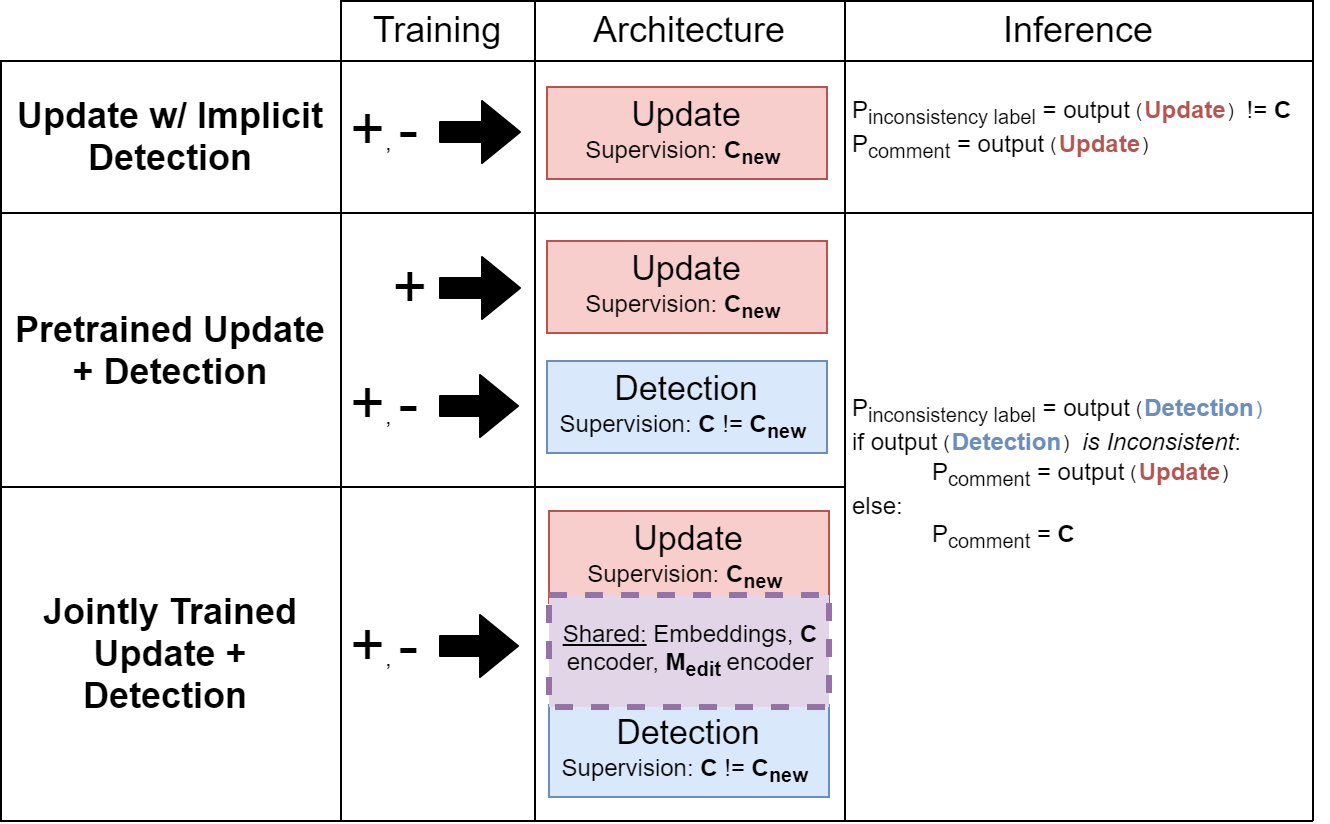}
\vspace{-5pt}
\caption{Overview of the three configurations used for the combined system which makes a prediction for whether \Comment{} is inconsistent ($P_{inconsistent}$) as well for the updated comment ($P_{comment}$). Modules are trained on either the full training set, including positive/inconsistent (+) and negative/consistent (-) examples, or only positive examples.}
\label{fig:combined_configs}
\end{figure*}

\begin{table*}
\centering
\small
\begin{tabular}{l@{\hskip 3mm}llllll@{\hskip 1mm}llll}
\hline
& \multicolumn{5}{c}{\bf Update Metrics} & & \multicolumn{4}{c}{\bf Detection Metrics} \\
\cline{2-6}
\cline{8-11}
& \bf xMatch & \bf METEOR & \bf BLEU-4 & \bf SARI & \bf GLEU & & \bf P & \bf R & \bf F1 & \bf Acc \\
\hline
Never Update & 50.0 & 67.7 & 71.6 & 25.1 & 68.3 & & 0.0 & 0.0 & 0.0 & 50.0 \\
\citeauthor{panthaplackel2020update}~\shortcite{panthaplackel2020update} & 21.5  & 56.2 & 64.7 & 37.6$\ssymbol{1}$ & 63.4 && 53.1 & 91.8 & 67.2 & 55.3 \\
\hline

\UpdateCopy{} & 56.1$\ssymbol{1}$ & 71.3 & 73.4$\ssymbol{1}$ & 30.2 & 71.4 && \bf 98.5 & 18.2 & 30.8 & 59.0\\
\hline

\Pretrained{} &  &  &  &  &  & &  &  &   &  \\
\hspace{0.2cm}\JustInTimeSeq{} + features &\bf  57.3$\ssymbol{4}$ & \bf 72.6$\ssymbol{1}$ & \bf 73.9$\ssymbol{2}$ & 37.8$\ssymbol{4}$ & \bf 73.2$\ssymbol{4}$ && 88.4$\ssymbol{2}$ & 73.2 & 80.0$\ssymbol{2}$ & 81.8$\ssymbol{1}$$\ssymbol{2}$  \\
\hspace{0.2cm}\JustInTimeGraph{} + features &  55.2 & 71.8 & 73.5$\ssymbol{1}$ & 38.0$\ssymbol{2}$$\ssymbol{6}$ & 72.8$\ssymbol{1}$ && 83.8 & \bf 78.3 & 80.9$\ssymbol{1}$ & 81.5$\ssymbol{2}$
\\
\hspace{0.2cm}\JustInTimeHybrid{} + features &  \bf 57.3$\ssymbol{4}$ & \bf 72.6$\ssymbol{1}$ & \bf 73.9$\ssymbol{2}$ & 37.6$\ssymbol{1}$ & \bf 73.2$\ssymbol{2}$$\ssymbol{4}$ && 88.6$\ssymbol{2}$ & 72.4 & 79.6$\ssymbol{2}$ & 81.5$\ssymbol{2}$	\\
\hline

\JointlyTrained{} &  &  &  &  &  & &  &  &   &  \\
\hspace{0.2cm}\JustInTimeSeq{} + features & 56.5$\ssymbol{1}$$\ssymbol{2}$ & 72.2$\ssymbol{2}$$\ssymbol{4}$ & 73.5$\ssymbol{1}$ & 37.9$\ssymbol{2}$$\ssymbol{6}$ & 72.9$\ssymbol{1}$ && 85.7$\ssymbol{1}$ & 76.7$\ssymbol{1}$ & 80.9$\ssymbol{1}$ & 81.9$\ssymbol{1}$$\ssymbol{2}$ \\
\hspace{0.2cm}\JustInTimeGraph{} + features &  56.2$\ssymbol{1}$ & 72.0$\ssymbol{4}$ & 73.6$\ssymbol{1}$ & 37.8$\ssymbol{2}$$\ssymbol{4}$ & 73.0$\ssymbol{1}$$\ssymbol{2}$ && 85.9$\ssymbol{1}$ & 76.7$\ssymbol{1}$ & \bf 81.0$\ssymbol{1}$ & 82.0$\ssymbol{1}$$\ssymbol{2}$ \\
\hspace{0.2cm}\JustInTimeHybrid{} + features &  56.8$\ssymbol{2}$ & 72.4$\ssymbol{1}$$\ssymbol{2}$ & 73.8 & \bf 38.1$\ssymbol{6}$ & 73.1$\ssymbol{4}$ && 86.7 & 75.7 & 80.9$\ssymbol{1}$ & \bf 82.1$\ssymbol{1}$ \\
\hline
\end{tabular}
\vspace{-5pt}
\caption{\label{table:full-comment-update-table} Results on joint inconsistency detection and update on the full test set. Scores for which the difference in performance is \textit{not} statistically significant are shown with identical symbols.}
\end{table*}

\begin{table*}[ht]
\centering
\small
\begin{tabular}{c@{\hskip 2mm}l@{\hskip 4mm}ccccccccc}
\hline
& & \multicolumn{4}{c}{\bf Cleaned Test Sample} & & \multicolumn{4}{c}{\bf Full Test} \\
\cline{3-6}
\cline{8-11}
& \bf Model & \bf P & \bf R & \bf F1 & \bf Acc & & \bf P & \bf R & \bf F1 & \bf Acc\\
\hline
\multirow{3}{*}{Post hoc}
& \PosthocSeq{}  & 58.9 & 68.0 & 63.0 & 60.3 & & 60.6 & 73.4 & 66.3 & 62.8\\
& \PosthocGraph{}{}  & 60.6 & 70.2 & 65.0 & 62.2 & & 62.6 & 72.6 & 67.2 & 64.6 \\
& \PosthocHybrid{}{}  & 53.7 & 77.3 & 63.3 & 55.2 & & 56.3 & \bf 80.8 & 66.3 & 58.9 \\
\hline
\multirow{3}{*}{Just-In-Time (implicit)} 
& \JustInTimeSeqImplicit{}   & 57.8 & 67.1 & 61.6 & 58.3 & & 61.5 & 74.0 & 66.9 & 63.4 \\
& \JustInTimeGraphImplicit{}  & 58.5 & 67.3 & 62.6 & 59.9 & & 61.3 & 71.8 & 66.1 & 63.3 \\
& \JustInTimeHybridImplicit{}  & 58.8 & 62.9 & 59.9 & 58.7 & & 62.9 & 69.3 & 65.1 & 63.1 \\

\hline
\multirow{3}{*}{Just-In-Time} 
& \JustInTimeSeq{}  & 83.8 & 79.3 & 81.5 & 82.0& & 80.7 & 73.8 & 77.1 & 78.0 \\
& \JustInTimeGraph{}  & 84.7 & 78.4 & 81.4 & 82.0 & & 79.8 & 74.4 & 76.9 & 77.6 \\
& \JustInTimeHybrid{}  & \bf 87.1 & \bf 79.6 & \bf 83.1 & \bf 83.8 & & \bf 80.9 & 74.7 & \bf 77.7& \bf 78.5 \\

\hline
\end{tabular}
\vspace{-5pt}
\caption{\label{table:implicit-explicit-table}Analyzing implicit code edit representations. Differences in F1 and Acc between \JustInTime{} (explicit) models vs. \Posthoc{} models and \JustInTime{} (explicit) vs. \JustInTime{} (implicit) are statistically significant.}
\end{table*}

 \begin{table*}[ht]
\centering
\small
\begin{tabular}{l@{\hskip 3mm}llllll@{\hskip 1mm}llll}
\hline
& \multicolumn{5}{c}{\bf Update Metrics} & & \multicolumn{4}{c}{\bf Detection Metrics} \\
\cline{2-6}
\cline{8-11}
& \bf xMatch & \bf METEOR & \bf BLEU-4 & \bf SARI & \bf GLEU & & \bf P & \bf R & \bf F1 & \bf Acc \\
\hline
Never Update  & 50.0 & 67.4 & 72.1$\ssymbol{5}$  & 24.9 & 68.2&  & 0.0 & 0.0 & 0.0 & 50.0 \\
\citeauthor{panthaplackel2020update}~\shortcite{panthaplackel2020update} & 25.9 & 60.0 & 68.7  & \bf 42.0 & 67.4$\ssymbol{4}$&  & 54.0 & \bf 95.6 & \bf 69.0 & 57.1$\ssymbol{4}$ \\
\hline
\UpdateCopy{} &  &  &  &  &  & &  & &  &  \\
\hspace{0.2cm}(\Comment{}, \NewCode{}) $\rightarrow$ \NewComment{} & 49.2 & 67.0 & 71.5$\ssymbol{1}$ &  27.9 & 68.4$\ssymbol{1}$$\ssymbol{2}$&  & 77.4 & 20.0$\ssymbol{1}$ & 31.4$\ssymbol{1}$ & 57.3$\ssymbol{4}$ \\
\hspace{0.2cm}(\Comment{}, \OldCode{}, \NewCode{}) $\rightarrow$ \NewComment{}  & 48.0 & 66.4 & 71.4$\ssymbol{1}$ &  25.4$\ssymbol{1}$ & 67.6$\ssymbol{4}$&  & 68.4$\ssymbol{5}$ & 8.7 & 15.4 & 52.3 \\
\hspace{0.2cm}(\Comment{}, \EditCode{}) $\rightarrow$ \NewComment{}  & 50.9$\ssymbol{1}$ & 68.1 &  72.1$\ssymbol{2}$$\ssymbol{4}$$\ssymbol{5}$ & 28.5 & 69.2&  & 80.1 & 20.2$\ssymbol{1}$ & 32.2$\ssymbol{1}$ & 57.6$\ssymbol{4}$ \\

\hspace{0.2cm}(\Comment{}, \NewCode{}) $\rightarrow$ \EditComment{}  $\Rightarrow$ \NewComment{} & 50.6$\ssymbol{1}$ & 67.7 & 72.3$\ssymbol{2}$ &  25.4$\ssymbol{1}$ & 68.6$\ssymbol{1}$&  & \bf 100.0$\ssymbol{4}$ & 1.8$\ssymbol{2}$ & 3.5$\ssymbol{2}$ & 50.9$\ssymbol{5}$ \\
\hspace{0.2cm}(\Comment{}, \OldCode{} \NewCode{}) $\rightarrow$ \EditComment{}  $\Rightarrow$ \NewComment{} & 50.3 & 67.5 & 72.2$\ssymbol{4}$ &  25.3 & 68.4$\ssymbol{2}$&  & 66.7$\ssymbol{5}$ & 1.6$\ssymbol{2}$ & 3.0$\ssymbol{2}$ & 50.8$\ssymbol{5}$ \\
\hspace{0.2cm}(\Comment{}, \EditCode{}) $\rightarrow$ \EditComment{}  $\Rightarrow$ \NewComment{}  & 52.1 & 68.6 & 72.9 & 26.9 & 69.6&  & \bf 100.0$\ssymbol{4}$ & 7.1 & 13.2 & 53.6 \\
\hspace{0.6cm}+features  & \bf 58.0 & \bf 72.0 & \bf 74.7 & 31.5 & \bf 72.7&  & \bf 100.0$\ssymbol{4}$ &  23.3 &  37.7 & \bf 61.7 \\



\hline


\end{tabular}
\vspace{-5pt}
\caption{\label{table:clean-update-configs}Results for various configurations of \UpdateCopy{} on the cleaned test sample. Scores for which the difference in performance is \textit{not} statistically significant are shown with identical symbols.}
\end{table*}
																																		
  \begin{table*}[t!]
\centering
\small
\begin{tabular}{l@{\hskip 3mm}llllll@{\hskip 1mm}llll}
\hline
& \multicolumn{5}{c}{\bf Update Metrics} & & \multicolumn{4}{c}{\bf Detection Metrics} \\
\cline{2-6}
\cline{8-11}
& \bf xMatch & \bf METEOR & \bf BLEU-4 & \bf SARI & \bf GLEU & & \bf P & \bf R & \bf F1 & \bf Acc \\
\hline

Never Update  & 50.0$\ssymbol{2}$ & 67.7$\ssymbol{4}$ & 71.6$\ssymbol{2}$ &  25.1 & 68.3&  & 0.0 & 0.0 & 0.0 & 50.0 \\
\citeauthor{panthaplackel2020update}~\shortcite{panthaplackel2020update} & 21.5 & 56.2 & 64.7 &  \bf 37.6 & 63.4&  & 53.1 & 91.8 & \bf 67.2 & 55.3$\ssymbol{1}$$\ssymbol{2}$$\ssymbol{4}$ \\
\hline
\UpdateCopy{} &  &  &  &  &  & &  & &  &  \\
\hspace{0.2cm}(\Comment{}, \NewCode{}) $\rightarrow$ \NewComment{} & 48.7 & 66.9 & 70.7$\ssymbol{1}$ &  27.1$\ssymbol{1}$ & 67.9&  & 73.8 & 16.9$\ssymbol{1}$ & 27.1 & 55.6$\ssymbol{1}$ \\
\hspace{0.2cm}(\Comment{}, \OldCode{}, \NewCode{}) $\rightarrow$ \NewComment{}  & 47.9 &  66.4 & 70.6$\ssymbol{1}$ & 25.6$\ssymbol{2}$ & 67.5&  & 65.2 & 8.7 & 15.3$\ssymbol{2}$ & 52.0 \\

\hspace{0.2cm}(\Comment{}, \EditCode{}) $\rightarrow$ \NewComment{}  & 50.0$\ssymbol{1}$$\ssymbol{2}$ & 67.7$\ssymbol{1}$$\ssymbol{2}$$\ssymbol{4}$ & 71.2 & 27.9 & 68.6$\ssymbol{1}$$\ssymbol{2}$&  & 78.7 & \bf 18.7$\ssymbol{2}$ & 30.1$\ssymbol{1}$ & 56.8$\ssymbol{4}$ \\

\hspace{0.2cm}(\Comment{}, \NewCode{}) $\rightarrow$ \EditComment{}  $\Rightarrow$ \NewComment{} & 50.2$\ssymbol{1}$ & 67.9$\ssymbol{1}$ & 71.7 &  25.6$\ssymbol{2}$ & 68.5$\ssymbol{1}$&  & \bf 100.0$\ssymbol{4}$ & 2.3 & 4.6 & 51.2 \\

\hspace{0.2cm}(\Comment{}, \OldCode{} \NewCode{}) $\rightarrow$ \EditComment{}  $\Rightarrow$ \NewComment{}  & 50.0$\ssymbol{2}$ & 67.8$\ssymbol{2}$ & 71.6$\ssymbol{2}$ &  25.2 & 68.4$\ssymbol{2}$&  & 93.3$\ssymbol{1}$$\ssymbol{4}$ & 0.8 & 1.5 & 50.4 \\

\hspace{0.2cm}(\Comment{}, \EditCode{}) $\rightarrow$ \EditComment{}  $\Rightarrow$ \NewComment{}  & 52.0 & 68.9 & 72.2 &  27.0$\ssymbol{1}$ & 69.4&  & 99.6$\ssymbol{4}$ & 7.4 & 13.7$\ssymbol{2}$ & 53.7$\ssymbol{2}$ \\
\hspace{0.6cm}+features & \bf 56.1 & \bf 71.3 & \bf 73.4 & 30.2 & \bf 71.4&  & 98.5$\ssymbol{1}$ & 18.2$\ssymbol{1}$$\ssymbol{2}$ & 30.8$\ssymbol{1}$ & \bf 59.0 \\


\hline
\end{tabular}
\vspace{-5pt}
\caption{\label{table:full-update-configs}Results for various configurations of \UpdateCopy{} on the full test set. Scores for which the difference in performance is \textit{not} statistically significant are shown with identical symbols.}
\vspace{-8pt}
\end{table*}

\section{Combined Detection+Update (Full)}
\label{appendix:combined}
We illustrate our approaches for combining the tasks of detection and update in Figure~\ref{fig:combined_configs}. In Table~\ref{table:full-comment-update-table}, we show results of combined detection+update systems on the full test set. The results are analogous to those presented in Section~\ref{sec:extrinsic-results} for the cleaned test set. While the differences for the update metrics are less pronounced, the pretrained and jointly trained approaches can again outperform \textit{\UpdateCopy{}} as well as the two reference points: Never Update and \citet{panthaplackel2020update}. The drastic differences in performance with respect to the detection metrics further demonstrate the importance of explicit inconsistency detection in a combined detection+update system. In line with our observations from the cleaned test set, we find the performances of the pretrained and jointly trained systems to be very close.

\section{Implicit vs. Explicit Edits}
\label{appendix:implicit_explicit}
Through \EditCode{} and \EditTree{}, we are \textit{explicitly} defining the code edits between \OldCode{} and \NewCode{} or between \OldTree{} and \NewTree{}. During preliminary experiments, we also considered having models \textit{implicitly} learn the edits. Namely, instead of providing \EditCode{} as the input to the sequence-based code encoder, we encode \OldCode{} and \NewCode{} separately. Both of these are encoded using the same GRU encoder, but multi-head attention is computed with the two sets of hidden states separately and then combined. We do the same for the graph-based approach (i.e., encode \OldTree{} and \NewTree{} separately rather than use \EditTree{}) as well as the hybrid approach.
Results for these approaches are shown in Table~\ref{table:implicit-explicit-table}, where \JustInTimeSeqImplicit{}, \JustInTimeGraphImplicit{},  and \JustInTimeHybridImplicit{} correspond to the encoding edits implicitly for the sequence-based, graph-based, and hybrid approaches implicitly. We find that implicitly encoding edits leads to performance that is similar (or even worse in some cases) than the \Posthoc{} setting. To truly take advantage of the \JustInTime{} setting, we find it necessary to encode edits explicitly, which can boost performance by wide, statistically significant margins. 

\section{Update w/ Implicit Detection Configurations}
 
 \begin{figure*}[t!]
\centering
\subfigure[Example from OpenAPI Generator]{
    \label{fig:openai}
    \includegraphics[scale=0.23]{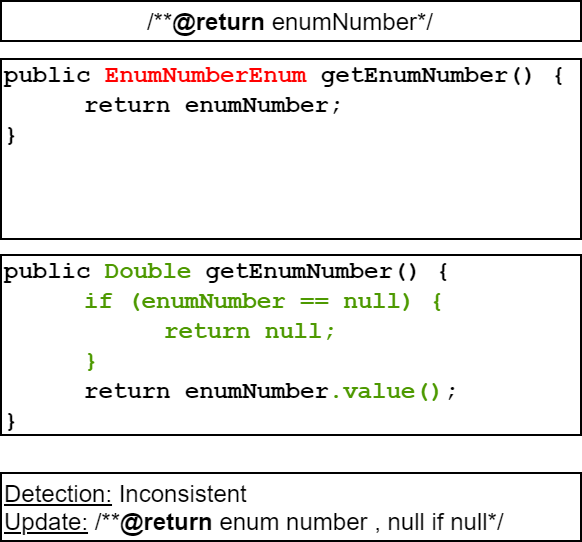}
}
\subfigure[Example from OWASP ZAP]{
    \label{fig:zaproxy}
        \includegraphics[scale=0.23]{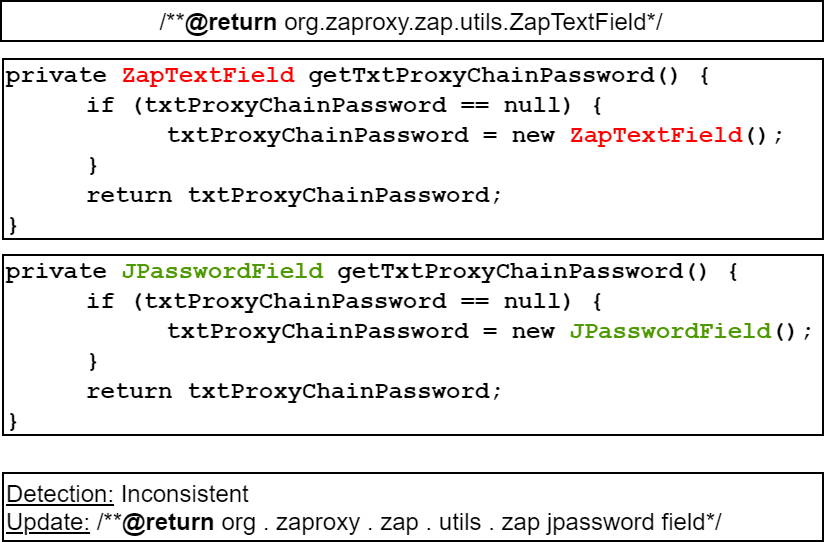}
}
\vspace{-5pt}
\caption{ Examples in which inconsistencies emerged as a result of developers failing to update comments upon code changes. Predictions of the combined, pre-trained detection+update approach are shown.}
\label{fig:real_example}
\end{figure*}

 The update components of our combined detection+update systems are based on the architecture proposed by ~\citet{panthaplackel2020update} for automatically updating comments based on code changes. As mentioned in Section~\ref{sec:extrinsic-method}, their approach entails encoding \Comment{} and a sequential code edit representation (\EditCode{}), and then using attention and a pointer network over these learned representations to decode a sequence of comment edit actions (\EditComment{}). The edit action sequence is finally parsed into an actual comment (\NewComment{}) as a post-processing step. This approach was initially designed to handle only cases in which a comment has to be updated. Our \textit{\UpdateCopy{}} baseline model applies this approach on both positive (i.e., inconsistent comments that should be updated) and negative (i.e., consistent comments that should not be updated) examples. Since their approach was not designed to support negative examples, we evaluate whether other input/output configurations of their architecture would be better suited for our setting. Because the features proposed in ~\citet{panthaplackel2020update} were tailored towards the specific inputs and outputs of their architecture, we disable features for these various configurations. However, we do evaluate how these configurations compare to the full model, which includes features.
 
 We present results for the cleaned test sample and full test set in Tables~\ref{table:clean-update-configs} and~\ref{table:full-update-configs}. Our notation for input/output configurations is as follows: \textit{(inputs) $\rightarrow$ output}. If the model generates \EditComment{} which is then parsed into \NewComment{}, we use the following notation on the output side: \textit{\EditComment{} $\Rightarrow$ \NewComment{}}. Note that \textit{(\Comment{}, \EditCode{}) $\rightarrow$ \EditComment{}  $\Rightarrow$ \NewComment{} + features} corresponds directly to training ~\citet{panthaplackel2020update}'s model on the full training set (positive and negative examples), i.e., the \textit{\UpdateCopy{}} model used in the main paper.

Similar to our findings from Appendix~\ref{appendix:implicit_explicit}, we observe that explicitly encoding code edits (\EditCode{}) significantly boosts performance across most metrics, independent of the output configuration. Training the model to generate \EditComment{} appears to yield improved performance across most update metrics, but this appears to deteriorate performance w.r.t. the detection metrics. However, none of these configurations that rely on implicit inconsistency detection can outperform \citet{panthaplackel2020update}'s approach (trained only on positive examples) on detection F1 or Acc. By incorporating features, we see improvements across most metrics, but the SARI and F1 metrics are still substantially higher for \citet{panthaplackel2020update}. Furthermore, the pretrained and jointly trained models presented in tables~\ref{table:comment-update-table} and \ref{table:full-comment-update-table} can outperform all of these configurations. This study confirms that even under various input/output configurations, without an explicit inconsistency detection component, a combined detection+update system cannot adequately identify inconsistent comments. Even in scenarios in which a possible update cannot be suggested, we argue that flagging inconsistencies and alerting developers would be a critical functionality of such a system. Therefore, implicitly performing inconsistency detection, as done by the various configurations of \textit{\UpdateCopy{}}, is not sufficient.
 
\section{Detecting Existing Inconsistencies}
 Recall that in our data collection procedure, we assign the negative (i.e., consistent) label to examples in which the developer did not update a comment following code changes. Based on our inspection of a sample of the full, unannotated test set, we find examples that are mislabeled as negative, and our model can correctly identify some of these cases. For instance, in the example shown in Figure~\ref{fig:openai}, the developer failed to amend the comment to indicate that the method no longer returns \CodeIn{enumNumber} but rather its value or \texttt{null} if it is not set. Similarly, in Figure~\ref{fig:zaproxy},  the developer failed to update \CodeIn{ZapTextField} to \CodeIn{JPasswordField} in the comment when the return type of the method was modified. The inconsistency in the OWASP ZAP project was fixed after we reported the issue and the inconsistency in the OpenAPI Generator project continues to persist today (at the time of submission). \footnote{We have reported them as issues in their respective projects. Warning: searching for these issues may reveal authors' identities.}
 
 Our inconsistency detection model correctly predicts the positive label for both of these cases, suggesting that it
 would have been able to potentially prevent these inconsistencies by alerting developers \JustInTime{}. Furthermore, by combining our pretrained \JustInTimeHybrid{} + features detection model with a pretrained update model~\cite{panthaplackel2020update}, we can additionally produce suggestions for resolving these inconsistencies. Note that because the update model is trained on subtokenized comments, it is not able to produce tokens like \texttt{enumNumber}; however, simple heuristics could be used to conjoin subtokens in the final prediction. While the suggested updates are not perfect, they could serve as starting points to help guide developers in updating comments. Nonetheless, this suggests that our approach can also be applied to detect existing inconsistencies (i.e., \Posthoc{} detection) by analyzing the history of changes to a particular comment/method pair.
 
  \begin{table*}[t!]
\centering
\small
\begin{tabular}{ll@{\hskip 3mm}ll}
\hline
& & \bf \# Epochs & \bf Training time\\
\hline
\multirow{9}{*}{Detection}
& \PosthocSeq{} & 16.3 & 1h 35.7s \\
& \PosthocGraph{} & 15.3 & 26m 16s    \\
& \PosthocHybrid{} & 28.3 & 2h 28m 10.3s \\
\cline{2-4}
& \JustInTimeSeq{} & 18.7 & 1h 23m 24.0s  \\
& \JustInTimeGraph{} & 15.0 & 29m 5.7s   \\
& \JustInTimeHybrid{} & 14.7 & 1h 31m 42.3s  \\
\cline{2-4}
& \JustInTimeSeq{} + features & 18.7 & 2h 38m 23s \\
& \JustInTimeGraph{} + features & 17.3 & 1h 5m 21.7s  \\
& \JustInTimeHybrid{} + features & 16.0 &  2h 51m 15.3s  \\
\hline
\multirow{15}{*}{Combined Detection+Update}
& \UpdateCopy{} & 33.7 & 43m 20.7s  \\
\cline{2-4}
& \Pretrained{} &  &    \\
& \hspace{0.2cm}\JustInTimeSeq{} + features & 57.0 & 3h 2m 59s \\
& \hspace{0.6cm} Detection & 18.7 & 2h 38m 23s  \\
& \hspace{0.6cm} Update & 38.3 & 24m 36s \\
& \hspace{0.2cm}\JustInTimeGraph{} + features & 55.6  & 1h 29m 57.7s  \\
& \hspace{0.6cm} Detection & 17.3 & 1h 5m 21.7s  \\
& \hspace{0.6cm} Update & 38.3 & 24m 36s  \\
& \hspace{0.2cm}\JustInTimeHybrid{} + features & 54.3 & 3h 15m 51.3s \\
& \hspace{0.6cm}  Detection &  16.0 &  2h 51m 15.3s   \\
& \hspace{0.6cm} Update & 38.3 & 24m 36s   \\
\cline{2-4}
& \JointlyTrained{} &  &    \\
& \hspace{0.2cm}\JustInTimeSeq{} + features & 30.7 &  45m 22s \\
& \hspace{0.2cm}\JustInTimeGraph{} + features & 25.7 & 45m 17s  \\
& \hspace{0.2cm}\JustInTimeHybrid{} + features & 27.3 & 46m 11s \\
\hline
\end{tabular}
\vspace{-5pt}
\caption{\label{table:inconsistency-detection-running-time} Average number of training epochs, and training time for inconsistency detection models as well as the combined detection+update models. Note that we used two different types of GPUs for these experiments, and therefore, the times are not necessarily comparable across models. Additionally, following every epoch of training the detection-only models, we compute precision, recall, F1, and Acc (using scikit-learn) on the validation data (as this determines the training termination condition), which adds to the computation time.}
\end{table*}

 \section{Hyperparameters}
Hyperparameters were tuned on validation data. For hidden dimension size, we considered \{64, 128\}. For number of attention heads, we considered \{1, 4, 8\}. For dropout, we considered \{0.2, 0.3, 0.6\}.

\section{Software and Hardware}
We implemented all neural models using PyTorch, and rely on PyTorch's default initialization methods for initializing model weights. We use the \textit{scikit-learn} library to compute evaluation metrics for inconsistency detection. All models were trained on a single GPU, either NVIDIA Titan V GPUs (12 GB) or GeForce GTX Titan Black GPUs (8 GB). The average number of training epochs as well as the average training times are provided in Table~\ref{table:inconsistency-detection-running-time}. 

\section{Statistical Significance}
We use bootstrap statistical significance testing~\cite{berg-kirkpatrick-etal-2012-empirical} with $p<0.05$ and 10,000 samples of size 5,000 each.

\end{document}